\documentclass[sigconf]{acmart}

\AtBeginDocument{%
  \providecommand\BibTeX{{%
    \normalfont B\kern-0.5em{\scshape i\kern-0.25em b}\kern-0.8em\TeX}}}

\copyrightyear{2023}
\acmYear{2023}
\setcopyright{acmlicensed}\acmConference[CCS '23]{Proceedings of the 2023 ACM SIGSAC Conference on Computer and Communications Security}{November 26--30, 2023}{Copenhagen, Denmark}
\acmBooktitle{Proceedings of the 2023 ACM SIGSAC Conference on Computer and Communications Security (CCS '23), November 26--30, 2023, Copenhagen, Denmark}
\acmPrice{15.00}
\acmDOI{10.1145/3576915.3623116}
\acmISBN{979-8-4007-0050-7/23/11}


  
\usepackage{multirow}
\usepackage{amsmath,amsthm,graphicx,mathtools,hyperref}
\usepackage{graphicx}
\usepackage{subfig}
\usepackage{textcomp}
\usepackage{xcolor}
\usepackage{xspace}
\usepackage{algorithm}
\usepackage{algorithmic}
\usepackage{balance}

\newcommand{\name}{OARS\xspace}
\newcommand{\longname}{Oracle-guided Adaptive Rejection Sampling\xspace}

 \hypersetup{
 colorlinks,
 linkcolor=blue
 }
\DeclareMathOperator*{\argmax}{arg\,max}

\definecolor{calpolypomonagreen}{rgb}{0.12, 0.7, 0.17}

\newcommand{\MYhref}[3][red]{\href{#2}{\color{#1}{#3}}}

\usepackage{stfloats}

\begin{document}

\title{Stateful Defenses for Machine Learning Models Are Not Yet Secure Against Black-box Attacks}

\author{\textbf{Ryan Feng}}
\thanks{$\dagger$ Denotes equal contribution.}
\email{rtfeng@umich.edu}
\authornotemark[2]
\affiliation{
  \institution{University of Michigan}
  \city{Ann Arbor}
  \state{Michigan}
  \country{USA}
}
\orcid{0000-0002-4767-274X}

\author{\textbf{Ashish Hooda}}
\authornotemark[2]
\email{ahooda@wisc.edu}
\affiliation{
  \institution{University of Wisconsin-Madison}
  \city{Madison}
  \state{Wisconsin}
  \country{USA}
}
\orcid{0000-0002-2928-919X}

\author{\textbf{Neal Mangaokar}}
\authornotemark[2]
\email{nealmgkr@umich.edu}
\affiliation{
  \institution{University of Michigan}
  \city{Ann Arbor}
  \state{Michigan}
  \country{USA}
}
\orcid{0000-0002-0684-4971}

\author{Kassem Fawaz}
\email{kfawaz@wisc.edu}
\affiliation{
  \institution{University of Wisconsin-Madison}
  \city{Madison}
  \state{Wisconsin}
  \country{USA}
}
\orcid{0000-0002-4609-7691}

\author{Somesh Jha}
\email{jha@cs.wisc.edu}
\affiliation{
  \institution{University of Wisconsin-Madison}
  \city{Madison}
  \state{Wisconsin}
  \country{USA}
}
\orcid{0000-0001-5877-0436}

\author{Atul Prakash}
\email{aprakash@umich.edu}
\affiliation{
  \institution{University of Michigan}
  \city{Ann Arbor}
  \state{Michigan}
  \country{USA}
}
\orcid{0000-0002-4907-3687}

\begin{abstract}
Recent work has proposed {\em stateful defense models (SDMs)} as a compelling strategy to defend against a black-box attacker who only has query access to the model, as is common for online machine learning platforms. Such stateful defenses aim to defend against black-box attacks by tracking the query history and detecting and rejecting queries that are ``similar'' and thus preventing black-box attacks from finding useful gradients and making progress towards finding adversarial attacks within a reasonable query budget. Recent SDMs (e.g., Blacklight and PIHA) have shown remarkable success in defending against state-of-the-art  black-box attacks. In this paper, we show that SDMs are highly vulnerable to a new class of adaptive black-box attacks. We propose a novel adaptive black-box attack strategy called \longname (\name) that involves two stages: (1) use initial query patterns to infer key properties about an SDM's defense; and, (2) leverage those extracted properties to design subsequent query patterns to evade the SDM's defense while making progress towards finding adversarial inputs. \name is broadly applicable as an enhancement to existing black-box attacks --  we show how to apply the strategy to enhance  six common black-box attacks  to be more effective against current class of SDMs. For example,  \name-enhanced versions of black-box attacks improved attack success rate against recent stateful defenses from almost 0\% to to almost 100\% for multiple datasets within reasonable query budgets. 
\end{abstract}

\maketitle

\section{Introduction}\label{sec:introduction}
Machine learning (ML) models are vulnerable to adversarial examples, imperceptibly modified inputs that a model misclassifies. Adversarial examples pose a significant threat to the deployment of ML models for applications such as deepfake detection~\cite{pu2021deepfake,hooda2022adversarially}, autonomous driving~\cite{geiger2012we, li2022blacklight}, medical image classification~\cite{gulshan2016development, wang2016deep}, or identity verification~\cite{sun2014deep}. Unfortunately, crafting defenses against white-box attackers who have full model access has proven difficult, particularly with the advent of adaptive attack strategies. There remains a large gap between natural and adversarial performance~\cite{athalye2018obfuscated,brown2017adversarial,croce2020reliable,eykholt2018robust,goodfellow2014explaining,madry2017towards,tramer2020adaptive}.

In recent years, there has been an increasing focus on a more restricted, but realistic black-box attack threat model, where an adversary only has query access to the model, e.g., via an API exposed by Machine Learning as a Service (MLaaS) platforms~\cite{amazonrekog,clarifai,plate_recognizer_2022}. The attacker can query the model for labels or label probability scores, but has no further access to the model or its training data. Several successful black-box attack methods have been proposed that use gradient or boundary estimation techniques to construct adversarial examples~\cite{brendel2017decision,chen2020hopskipjumpattack,feng2022graphite,ilyas2018black,li2020qeba,maho2021surfree,moon2019parsimonious,wierstra2014natural,yan2021policy}. However, such techniques typically require querying multiple nearby inputs to approximate the local loss landscape.

This observation has led to a new line of defense work we refer to as \emph{Stateful Defense Models (SDMs)}~\cite{chen2020stateful,li2022blacklight,choi2023piha,juuti2019prada}, which target black-box query-based attacks. Observing that such attacks sample multiple nearby points, SDMs use an internal state to track past queries. SDMs then monitor future queries and perform some restrictive action against the attacker when receiving queries that are too similar, where similarity is defined through some defense-chosen measure. Such an event is also referred to as a {\em collision}. Defensive actions can then include banning the user's account or rejecting queries. Blacklight~\cite{li2022blacklight} is a state-of-the-art SDM, which uses probabilistic content fingerprints to reject highly similar queries, thereby thwarting black-box attacks.

Since SDMs directly target the black-box attacks' fundamental reliance on issuing similar queries, attacking SDMs remains an open and challenging problem. Most contemporary black-box query-based attacks involve some combination of gradient estimation from averaging samples drawn from a distribution and walking along the decision boundary in small steps, both of which can easily result in account bans or query rejections (Section~\ref{sec:bb_elements}). The main method for adapting black-box attacks is query-blinding~\cite{chen2020stateful}, which applies simple transformations on input queries. It aims to make small changes to avoid detection without disrupting the attack optimization process. Blacklight~\cite{li2022blacklight}, in particular, have made remarkable progress, with a perfect 0\% attack success rate against a wide range of attacks, even with query-blinding (an observation we confirm in Section~\ref{sec:experiments}).

However, a related question naturally arises: Similarly to the white-box setting, where robustness was overstated due to a lack of adaptive attack evaluation~\cite{athalye2018obfuscated}, can SDMs be broken by stronger adaptive black-box attacks that attempt to evade query collisions?

In this paper, we find that SDMs are highly vulnerable to a new class of adaptive black-box attacks. \emph{The key insight underlying these attacks is that SDMs leak information about their similarity-detection procedure.} We use this information to adapt and enhance black-box attacks to be more effective against these SDMs. Our novel adaptive black-box attack strategy called \longname (\name) involves two stages: (1) use initial query patterns to extract information about the SDM's similarity-detection procedure; and (2) leverage this knowledge to design subsequent query patterns that evade the SDM's similarity check while making progress towards finding adversarial inputs (Section~\ref{sec:approach}). 

Using \name to work with a black-box attack is a non-trivial design problem -- multiple elements in a typical black-box attack must be modified to achieve the following tasks, while \textit{avoiding collisions}: (1) sampling to estimate gradients; (2) choosing an appropriate step size;  and (3) modifying the technique for finding the decision boundary in hard-label black-box settings. We show how \name applies such modifications in a principled way. First, we query the model to fine-tune a proposal probability distribution of the relevant parameters to each task. Second, we repeatedly sample the fine-tuned distribution to generate examples and use the defense as an oracle to reject or accept them. The accepted examples evade collisions and lead to more successful attacks (Section~\ref{sec:bb_elements}). 

We demonstrate how \name can broadly enhance a wide range of black-box attacks to make them more potent against SDMs.
In particular, we apply \name to six commonly used black-box query-based attacks (NES~\cite{ilyas2018black}, HSJA~\cite{chen2020hopskipjumpattack}, QEBA~\cite{li2020qeba}, SurFree~\cite{maho2021surfree}, Square~\cite{andriushchenko2020square}, Boundary~\cite{brendel2017decision}) (Section~\ref{sec:att_existing}). Through comprehensive empirical evaluation, we find that \name-enhanced versions of these black-box attacks significantly outperform both the standard versions and query-blinding on four contemporary stateful defenses (the original stateful detection defense~\cite{chen2020stateful}, Blacklight~\cite{li2022blacklight}, PIHA~\cite{choi2023piha}, and IIoT-SDA~\cite{esmaeili2022iiot}). For the best \name-enhanced black-box attacks, the attack success rate increased from close to 0\% (for the best defense) to almost 100\% for each of the four stateful defenses on multiple datasets with reasonable query budgets for the attacks to be practical (Section~\ref{sec:eval_vanilla}). Finally, we discuss potential directions for improving stateful defenses to counteract our new attacks. We tested several variations of existing SDMs to counteract our attacks but found that \name-enhanced black-box attacks maintained a high attack success rate (Section~\ref{sec:reconfigs_sdm}). 

In summary, this paper demonstrates that recent SDMs, thought to be strong defenses against black-box attacks, are actually highly vulnerable. We propose a technique called \longname (\name) that helps make multiple black-box attack algorithms much more potent against these SDMs. These \name-enhanced black-box attack methods thus provide a new benchmark to test any future proposed stateful defenses against black-box attacks.
\section{Background and Related Work}\label{sec:framework}
In this section, we describe our notation and terminology, threat model, black-box attacks, stateful defense models (SDMs), and the difficulties of attacking stateful defense models.

\subsection{Notation and Terminology}\label{sec:notation}
First, we introduce our common notation and terminology to describe the SDMs and the related attacks. 

\subsubsection{\textbf{General Notation}}
Let $\mathcal{D}$ be the distribution of input space $ \mathcal{X} \times \mathcal{Y}$, where $\mathcal{X} \in \mathbb{R}^d$ is the space of $d-$dimensional samples, and $\mathcal{Y}$ is the class label space. Let $F: \mathcal{X} \rightarrow [0,1]^{|\mathcal{Y}|}$ be the ``soft-label'' DNN classifier trained using loss function $L$ that outputs probabilities over the classes, i.e., $F(x)_i$ is the probability that $x \in \mathcal{X}$ belongs to the $i^{th}$ class. The ``hard-label'' classifier can then be given by $f(x) = \argmax_i ~F(x)_i$. Given a victim sample $x_{vic} \in \mathcal{X}$ with true label $y \in \mathcal{Y}$, and a perturbation budget $\epsilon$, an adversarial example is any sample $x_{adv}$ such that $f(x_{adv}) \neq y$ and $||x_{adv} - x_{vic}|| \leq \epsilon$ for an appropriate choice of norm. We describe an identity matrix in $\mathbb{R}^d$ by $I_d$.

For iterative black-box attacks, let $x_t$ represent the sample at the current $t^{th}$ iteration of the attack.

\subsubsection{\textbf{SDM Terminology}}
\label{sec:sdm-terminology}
To establish a standard setup for SDMs, we introduce the six components that make up a typical SDM. We then describe the SDMs we evaluate in terms of these named components in Section~\ref{sec:bg_stateful_defense_model}.

\paragraph{\textbf{Classifier.}}{
The classifier $F$ is the underlying model to be protected by the SDM.}

\paragraph{\textbf{Feature Extractor.}}{
The feature extractor is a function $h(\cdot)$ that extracts a representation of incoming queries to determine query similarity. The similarity of two samples $x_1$ and $x_2$ is the distance between $h(x_1)$ and $h(x_2)$.}

\paragraph{\textbf{Query Store.}}{ The query store is a finite-capacity stateful buffer $Q$ that stores the history of past queries. }

\paragraph{\textbf{Similarity Procedure.}}{ The similarity procedure $s$ uses $h(\cdot)$ to compute the distance of a given query to the most similar element(s) in the store.} 

\paragraph{\textbf{Action Function.}}{ The action function \verb|action| identifies collisions and takes the necessary steps to thwart the attack. Informally, a collision occurs when the similarity procedure determines the incoming query to be ``too similar'' to a previous query.} We consider SDMs with two different assumptions about user accounts (see Section~\ref{sec:bg_stateful_defense_model}). Depending on the underlying assumptions of the attackers, SDMs either act by banning the accounts~\cite{chen2020stateful} or rejecting queries~\cite{li2022blacklight,choi2023piha,esmaeili2022iiot} to limit the attacker.

\begin{figure}
    \centering
    \includegraphics[width=\columnwidth]{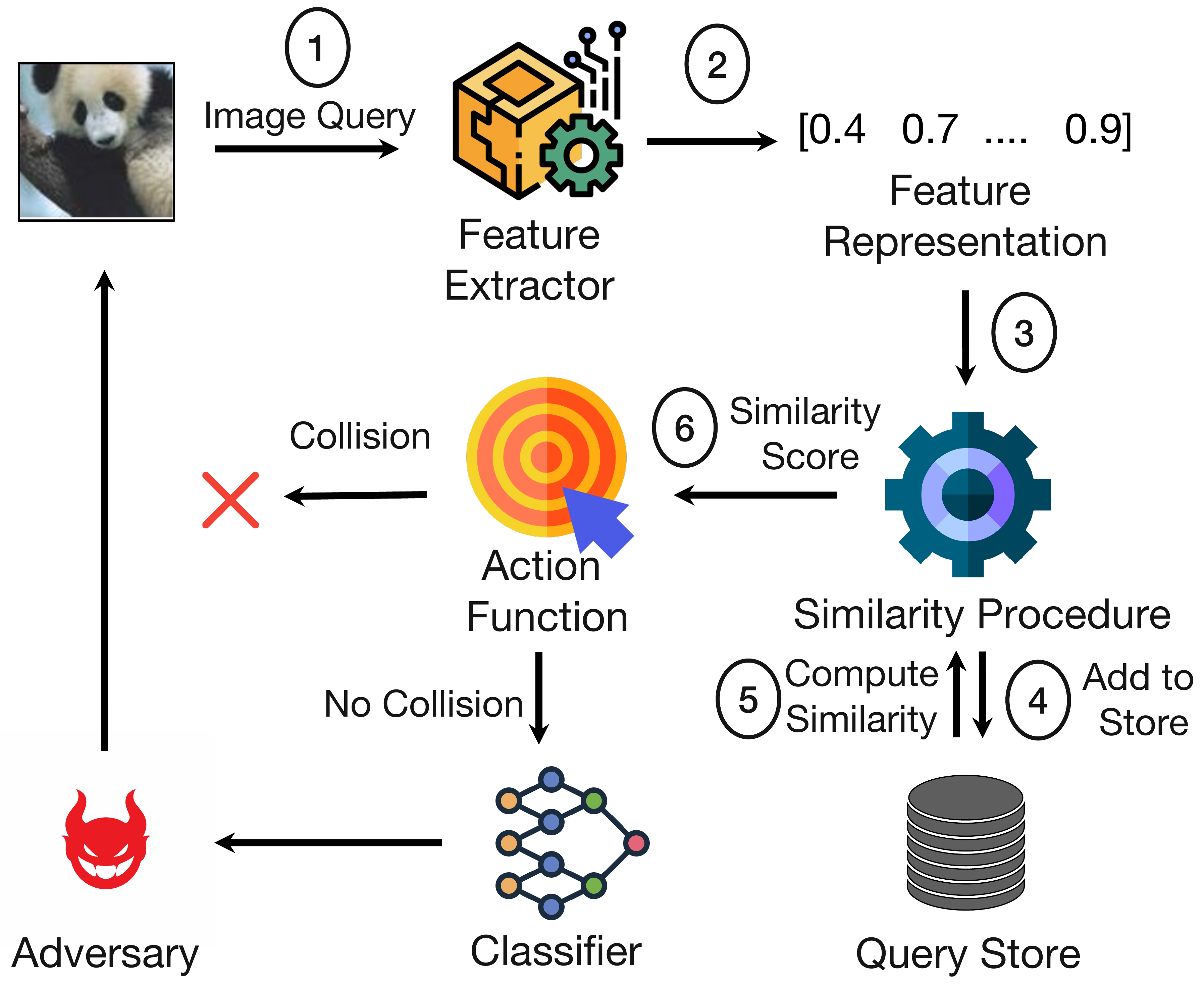}
    \caption{A general stateful defense pipeline. Given an input query, the input's features are compared for similarity with features of queries stored in the query store, resulting in a similarity score. It also adds the query to the query store. The action function then decides whether the similarity score indicates a collision. In case of no collision, the model is queried, and the model output is returned; else, an action is taken, such as rejecting the query.}
    \label{fig:stateful_framework_workflow}
\end{figure}

Figure~\ref{fig:stateful_framework_workflow} shows a conceptual pipeline of SDMs in terms of the above elements. Given an input query (for example, an image to be classified), the input sample passes through a feature extractor. The output feature representation is then passed to the similarity procedure, which compares this representation with those in the query store and outputs a similarity score. It also adds the query to the query store. The action function then decides whether the similarity score indicates a collision or not (a collision implies the query is similar to a stored query). If there is no collision, the model is queried, and the model output is returned. If there is a collision, typical action is to reject the query or ban a user's account.

\subsection{Threat Model}
We consider a black-box threat model where a queryable Machine-Learning-as-a-Service (MLaaS) platform hosts a classifier. Examples of such platforms include Clarifai~\cite{clarifai}, Amazon Rekognition~\cite{amazonrekog}, and Automatic License Plate Recognition systems~\cite{plate_recognizer_2022}. Users can submit queries by registering an account with the service and interacting with its public API. We focus on query-based attacks, where an attacker can only query the model for outputs, which can either be just the final label (hard-label) or include class probabilities (soft-label or score-based). We evaluate our SDMs against both score-based and hard-label query-based attacks~\cite{ilyas2018black,chen2020hopskipjumpattack,andriushchenko2020square,brendel2017decision,li2020qeba,maho2021surfree}.  \emph{The proposed \name attacks are model agnostic - we do not utilize any knowledge about the underlying model specifics, whether an SDM is deployed at all, or which specific SDM method if deployed, is being used}. 

If an SDM is deployed, we do not disable or pause the SDM for any stage of the attack in any of our experiments. This means that SDMs run unmodified for the entire duration of the attack and are free to take action by rejecting queries or banning accounts during all stages of our attacks, matching a practical deployment where the SDM will always be active. Our key insight is that \name can leverage these rejections or bans to adapt the attacks.

\subsection{Black-box Attacks}\label{sec:bb_attacks}

We analyze a set of black-box attacks this paper: NES~\cite{ilyas2018black}, HSJA~\cite{chen2020hopskipjumpattack}, QEBA~\cite{li2020qeba}, Boundary~\cite{brendel2017decision}, Square~\cite{andriushchenko2020square}, and SurFree~\cite{maho2021surfree}, which we briefly describe below. Following prior work~\cite{li2022blacklight}, we use $\ell_\infty$ versions of NES~\cite{ilyas2018black} and Square~\cite{andriushchenko2020square} and $\ell_2$ versions of the remaining attacks.

\paragraph{\textbf{NES (Score-based)~\cite{ilyas2018black}.}}{ NES is a score-based attack that aims to construct adversarial examples by estimating the loss gradient. NES uses finite differences over samples from a Gaussian distribution to estimate the gradient of the loss. It then performs projected gradient descent with the estimated gradient.}

\paragraph{\textbf{Square (Score-based)~\cite{andriushchenko2020square}.}}{
Square attack is a score-based attack that applies $\ell_p$-bounded random perturbations to pixels within the squares inside the input sample. It chooses progressively smaller squares to ensure attack success. }

\paragraph{\textbf{HSJA (Hard-label)~\cite{chen2020hopskipjumpattack}.} }{HopSkipJumpAttack (HSJA) is a hard-label attack that repeatedly (1) locates the model decision boundary via binary search, (2) estimates the gradient around the boundary via estimated local loss differences, and (3) identifies the optimal step size via geometric progression before performing gradient descent. }

\paragraph{\textbf{QEBA (Hard-label)~\cite{li2020qeba}.}}{ QEBA is a hard-label attack based on HopSkipJumpAttack that samples noise from a lower-dimensional subspace to estimate the gradient around the boundary.}

\paragraph{\textbf{SurFree (Hard-label)~\cite{maho2021surfree}.}}{ SurFree is a hard-label attack that avoids gradient estimation. Guided by the geometric properties of the decision boundary, it performs a random search of directions starting from a given sample to identify the direction closest to the decision boundary. 

\paragraph{\textbf{Boundary (Hard-label)~\cite{brendel2017decision}.}}{
Boundary is a hard-label attack that starts from a randomly initialized adversarial example and performs a ``boundary walk'' to reduce the perturbation size.}

\subsection{Stateful Defense Models}\label{sec:bg_stateful_defense_model}
 The key intuition behind stateful defense models is that query-based black-box attacks need to query ``similar'' samples. Thus, rejecting any query that is ``too similar'' to a previously queried sample, as described in Section~\ref{sec:sdm-terminology}, can help prevent these attacks from estimating accurate gradients. For example, NES~\cite{ilyas2018black} samples multiple Gaussian perturbed versions of a given sample $x_t$ to estimate the loss gradient, with such queries likely being similar to the original query $x_t$.
\begin{table*}[t]
\centering
     \caption{Existing defenses (OSD~\cite{chen2020stateful}, Blacklight~\cite{li2022blacklight}, PIHA~\cite{choi2023piha}, and IIoT-SDA~\cite{esmaeili2022iiot}) summarized in terms of their choices for each component.}
     \label{tab:instantiating_existing_defenses}
\begin{tabular}{ l c c c c}
\toprule
\textbf{Defense} & \textbf{Query Store} & \textbf{Feature Extractor} & \textbf{Similarity Procedure}        & \textbf{Action} \\ \midrule
OSD~\cite{chen2020stateful}              & Per Account          & Neural Encoder             & $k$NN over $\ell_2$ Distance ($k$ = 50) & Ban Account     \\
Blacklight~\cite{li2022blacklight}       & Global               & Pixel-SHA                  & $k$NN over Hamming Distance ($k$ = 1) & Reject Query    \\
PIHA~\cite{choi2023piha}            & Global               & PIHA's Percept. Hash       & $k$NN over Hamming Distance ($k$ = 1) & Reject Query    \\
IIoT-SDA~\cite{esmaeili2022iiot}         & Per Account          & Neural Encoder             & $k$NN over $\ell_2$ Distance ($k$ = 11) & Reject Query    \\ \bottomrule
\end{tabular}

\end{table*}

We now describe four existing SDMs (OSD~\cite{chen2020stateful}, Blacklight~\cite{li2022blacklight}, PIHA~\cite{choi2023piha}, and IIoT-SDA~\cite{esmaeili2022iiot}) using the terminology in Section~\ref{sec:sdm-terminology}. Table~\ref{tab:instantiating_existing_defenses} compares these SDMs. 

\paragraph{\textbf{OSD~\cite{chen2020stateful}.} }{Originally proposed by Chen et al.~\cite{chen2020stateful}, OSD employs a neural similarity encoder as the feature extractor $h$. The encoder outputs a $d-$dimensional embedding vector and is trained using contrastive loss~\cite{bell2015learning} to learn perceptual similarity. To detect whether incoming query $x$ induces a collision, OSD uses a per-account query store and similarity procedure that computes the average $\ell_2$ distance between $h(x)$ and its $k-$nearest neighbors in that account's query store. Finally, OSD's action function bans the user's account if a collision is detected. As Sybil accounts~\cite{douceur2002sybil,yang2014uncovering} successfully render the cost of multiple accounts minimal, OSD can be trivially broken. Blacklight~\cite{li2022blacklight} and PIHA~\cite{choi2023piha} address this limitation.}

\paragraph{\textbf{Blacklight~\cite{li2022blacklight}.}}{ 

Blacklight~\cite{li2022blacklight} is an SDM that differs from OSD~\cite{chen2020stateful} in two ways. First, it changes the feature extractor $h$ to a new probabilistic hash function, Pixel-SHA. Pixel-SHA quantizes pixel values of the input sample, hashes multiple segments of pixels at a given stride length,
and concatenates the top 50  hashes to form the final output. Since the output of $h$ is a hash, Blacklight employs an approximate nearest-neighbor search over normalized Hamming distances (which range between 0 and 1) as its similarity procedure, which can be computed in $O(1)$ time.

Second, with the assumption that Sybil accounts~\cite{douceur2002sybil,yang2014uncovering} can be easily created, Blacklight~\cite{li2022blacklight} employs a global hash-table as its query store. Since the store is global, collisions are detected regardless of the number of accounts the attacker creates. As a result of this change, Blacklight's action function chooses to reject queries, i.e., deny classification service when a collision is detected, instead of banning accounts.

Blacklight assumes that the global store is large but finite and needs to be reset infrequently once a day to handle the query workload. With a reset of once per day, Blacklight authors report that it would take an attacker approximately 3 years to complete the fastest successful attack if they waited for the store to reset before proceeding, which is a significant slowdown.}

\paragraph{\textbf{PIHA~\cite{choi2023piha}.}}{
PIHA~\cite{choi2023piha} is an image-specific SDM similar to Blacklight~\cite{li2022blacklight} in that it uses a hash function to detect similarities and uses a global store to detect collisions. PIHA, however, proposes a novel perceptual hashing algorithm to act as a feature extractor. This algorithm applies a low-pass filter, converts the image into the HSV or YC${}_{B}$C${}_{R}$ color spaces, and uses the Local Binary Patterns algorithm~\cite{ojala1996comparative} to compute the hash function on blocks of the input.}

\paragraph{\textbf{IIoT-SDA~\cite{esmaeili2022iiot}.}}{
IIoT-SDA~\cite{esmaeili2022iiot} is a defense targeting malware classification in an industrial IoT setting. IIoT-SDA reshapes executable byte codes into a matrix and trains a neural encoder with a Mahalanobis~\cite{mahalanobis1936generalized} defined contrastive loss function as its feature extractor $h$. IIoT-SDA uses a per-account query store and $k$-nearest neighbors as its similarity procedure, like OSD~\cite{chen2020stateful}. Like Blacklight~\cite{li2022blacklight}, IIoT-SDA's action function is to reject queries. Thus, IIoT-SDA can be viewed as a combination of OSD and Blacklight, adapted to the malware classification domain.}

\subsection{Query-blinding Attacks}\label{sec:query-blind}
SDMs have shown promise against black-box query-based attacks. For example, Blacklight~\cite{li2022blacklight} reports 0\% attack success rate for all of the black-box attacks in Section~\ref{sec:bb_attacks}. We attribute this success to attacks issuing similar queries to the target model. To avoid issuing similar queries,  Chen et al. proposed \emph{Query-blinding attacks}~\cite{chen2020stateful} that attempt to evade SDM detection by transforming samples before performing queries. The intuition is that it may be possible to transform samples so that the samples are no longer ``similar'' enough to be detected by an SDM but useful enough to infer gradients and enable the underlying optimization process to succeed. 

Formally, query-blinding is defined by two functions: a randomized blinding function $b(x; s)$ that maps the input query $x$ to a modified example $x'$ such that $x$ and $x'$ do not result in a collision, and a revealing function $r(F(x'))$ that estimates $F(x)$ from the blinding function and the classifier outputs ($F(x')$). For the image domain, query-blinding is most commonly implemented by taking image transformations (e.g., rotation, translation, scaling, Gaussian noise) and selecting a random value for each query within a range of values parameterized by $s$.

Still, Blacklight~\cite{li2022blacklight} and recent SDMs~\cite{chen2020stateful,choi2023piha} claim to defeat query-blinding attacks, suggesting that attacking SDMs remains a difficult and unsolved problem. In particular, query-blinding struggles with balancing the trade-off between attack utility and detection: more destructive transforms increase the odds of evading detection but also hurt the optimization process. We provide further experimental insight into why query-blinding fails in Section~\ref{sec:experiments}.
\section{Designing Adaptive Black-box Attacks}
\label{sec:approach}

We propose a novel adaptive black-box attack strategy against SDMs based on the following insight: \emph{the SDM's action module ``leaks'' information about its similarity procedure and query store.} An attacker can leverage information from the defender's past action behavior to issue queries that evade collisions while performing the attack. More formally, we represent the SDM as an oracle that enables the attacker to perform {\em rejection sampling} (discussed further in Section~\ref{sec:bb_elements}) to help future queries avoid collisions. This insight guides our proposed Oracle-guided Adaptive Rejection Sampling (\name) strategy, which introduces a two-pronged \emph{adapt} and \emph{resample} strategy to evade collisions.

We first describe three key elements of black-box attacks and how the attacker adapts each element using rejection sampling (Section~\ref{sec:bb_elements}). Then, we apply our general adaptive strategy to enhance six popular black-box attacks (NES~\cite{ilyas2018black}, HSJA~\cite{chen2020hopskipjumpattack}, QEBA~\cite{li2020qeba}, SurFree~\cite{maho2021surfree}, Square~\cite{andriushchenko2020square}, and Boundary~\cite{brendel2017decision}) (Section~\ref{sec:att_existing}).

\begin{figure}
     \centering
       \subfloat[\emph{\name-NES: Adapting the proposal distribution}. For gradient estimation, we adapt a Gaussian proposal distribution to estimate the optimal $\sigma_{opt}$ that achieves the target collision rate.\label{fig:hsja_ge_adapt}]{\includegraphics[width=1.25in]{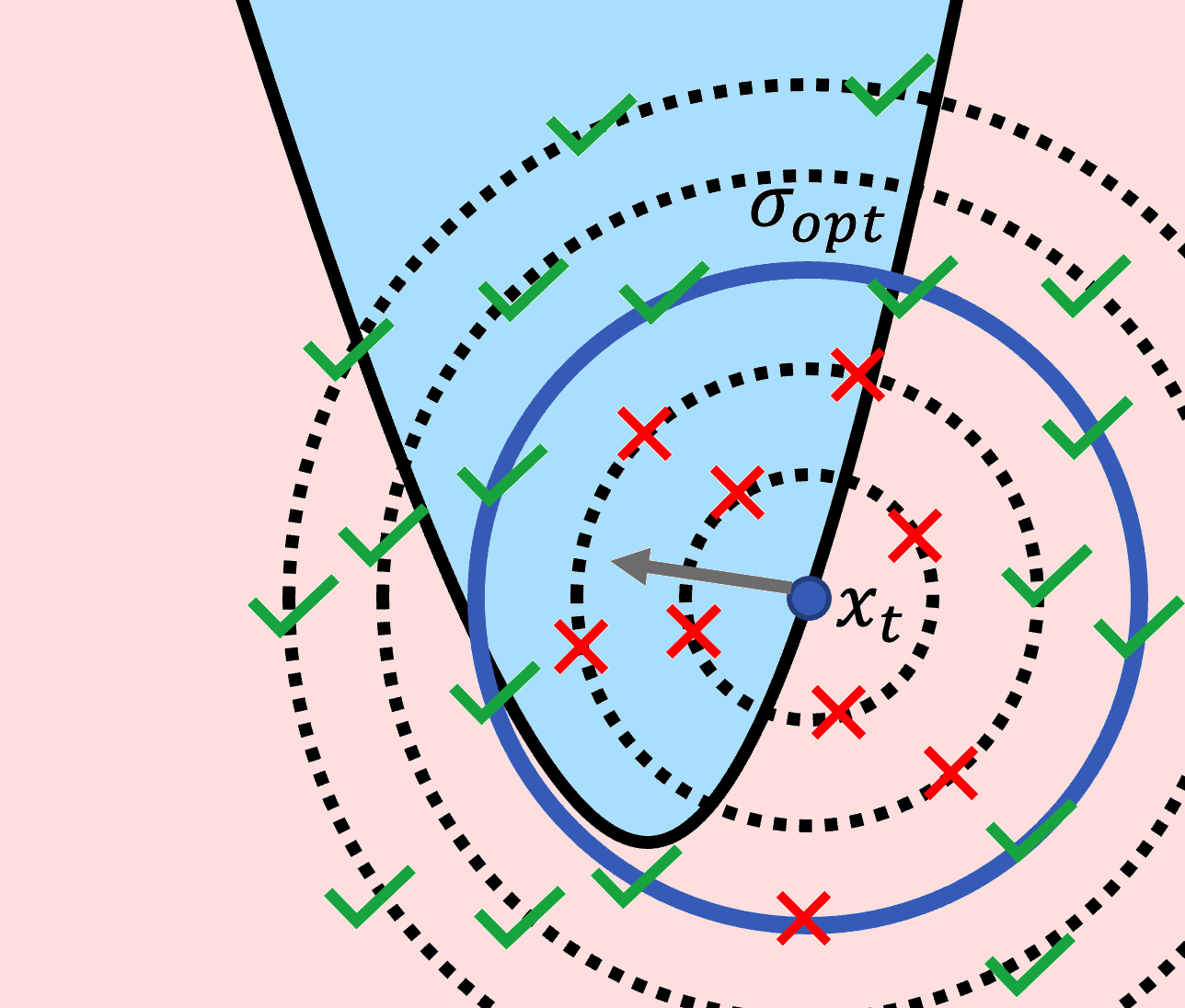}}
       \hfil
       \subfloat[\emph{\name-NES: Resampling}. For gradient estimation, once we have $\sigma_{opt}$, we perform rejection sampling over from $\mathcal{N}(0,\sigma_{opt}^2I_d)$. We resample up to $ge_{tries}$ to get $n$ valid samples and average over the valid samples we obtain. \label{fig:hsja_ge_using_proposal}]{\includegraphics[width=1.25in]{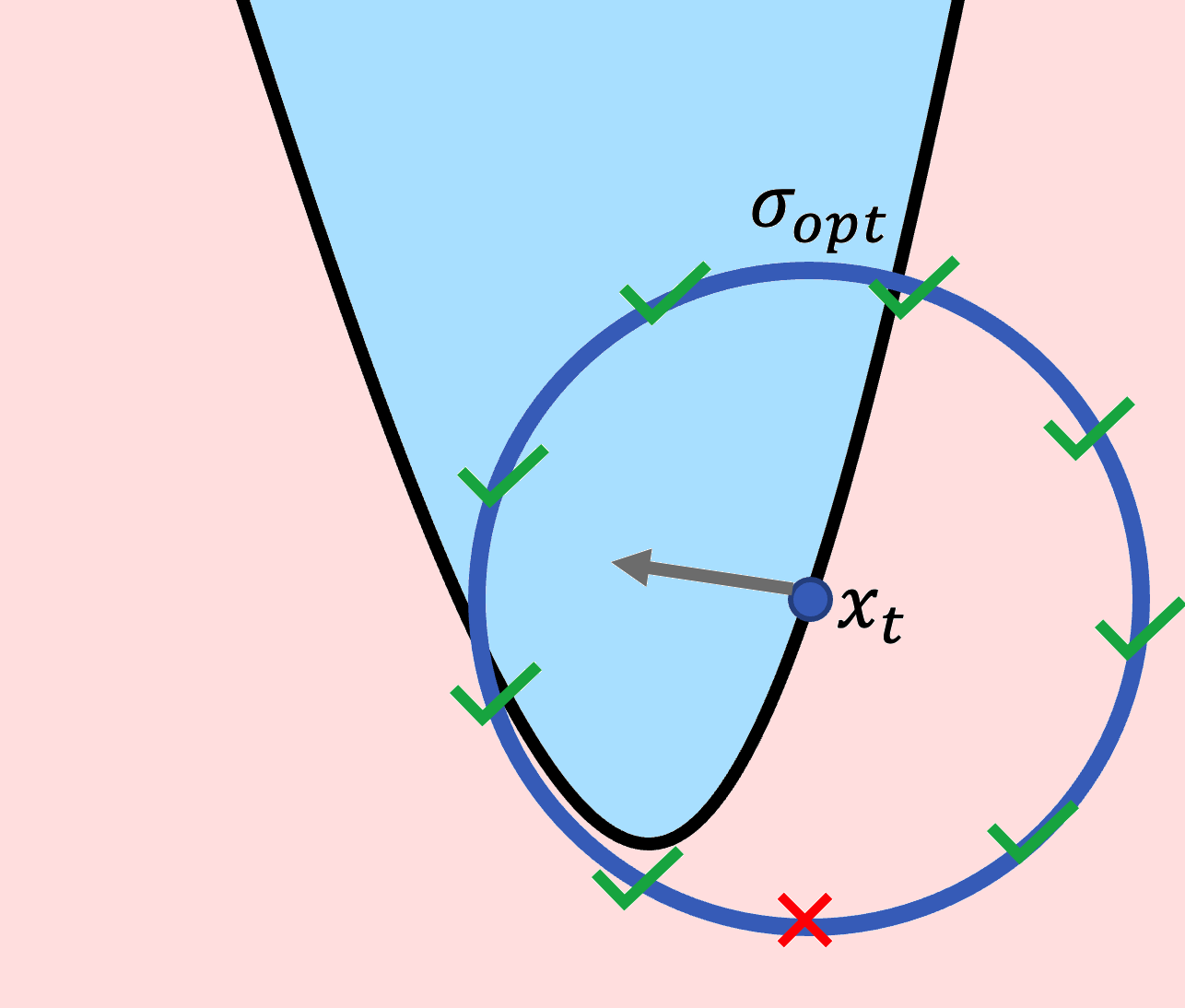}}
    \caption{Example illustration of the \name adapt and resample gradient estimation for NES~\cite{ilyas2018black}. Red X's denote queries that collide, and green checkmarks denote queries that do not collide. }\label{fig:examples}
\end{figure}

\subsection{Modifying Common Attack Elements with Oracle-Guided Adaptive Rejection Sampling}
\label{sec:bb_elements}

We begin with the following observation: Typical query-based attacks in the black-box settings involve three attack elements: estimating a gradient, taking a step, and locating the boundary. 

\begin{itemize}
    \item \textbf{Estimating a gradient:} Given a current attack point $x_t$, gradient estimation typically adds Gaussian noise scaled by a standard deviation of $\sigma$ and compares the differences in model outputs.
    
    \item \textbf{Taking a step:} After estimating the gradient, attackers update $x_t$ by taking a step in that direction. Black-box attacks typically anneal their step size to small values for improving convergence.
    
    \item \textbf{Locating the boundary:} After taking a step, many black-box attacks update the example by interpolating the line between the post-step attack point $x_t$ and the original sample $x_{vic}$. This interpolation moves the $x_t$ towards the boundary to select a better starting point for the next iteration. Typically, this takes the form of a binary search between $x_t$ and $x_{vic}$. 
\end{itemize}

An effective adaptive attack against SDMs must revisit these elements to issue queries that evade collisions while converging to a successful adversarial example. Towards that end, we propose a general strategy, Oracle-guided Adaptive Rejection Sampling (\name), which leverages the defense to adapt attack queries. Recall that SDMs thwart query-based black-box attacks by rejecting  ``similar'' queries. \name uses the SDM to select the most similar queries that will not collide with a given sample $x$. This way, it uses rejection sampling to pick queries from a target distribution $p^\prime_{x}(q)$, which represents the most similar potential queries that won't cause a collision with the sample $x$.
More specifically, rejection sampling selects a tractable \textit{proposal distribution} $p_x^{\theta}(q)$ as an estimate of $p^\prime_{x}(q)$, and then samples a query from $p_x^{\theta}(q)$. If this query collides, it can be discarded, and a new sample be drawn. 

However, selecting a good proposal distribution, i.e., selecting a good $\theta$ poses a key challenge --- sampling from a poorly selected proposal distribution will yield a large number of samples that do not belong to $p^\prime_{x}(q)$, i.e., submitting these queries causes frequent collisions and exhausts the query budget. To address this challenge, \name initializes a parametric proposal distribution $p_x^{\theta}(q)$ and uses the SDM as an oracle to find an estimated $\theta_{opt}$ such that $p_x^{\theta_{opt}}(q) \approx p^\prime_{x}(q)$. We assume that $\theta$ is a unimodal and monotonic parameter, i.e., the average collision rate decreases with an increase in $\theta$. Adaptation can be achieved via any suitable optimization algorithm. For example, an adversary can use the binary-search protocol in Algorithm~\ref{alg:adapt_proposal_grad_est} to fine-tune their $p_x^{\theta}(q)$. 

\begin{algorithm}[t]
\caption{\texttt{ADAPT\_PROPOSAL}: Fine-tune a parametric proposal distribution $\mathcal{N}(0, \sigma^2I_d)$ via Oracle-guided binary search for the gradient estimation attack element.}
\begin{algorithmic}[1]
\REQUIRE Proposal Distribution $\mathcal{N}(0, \sigma^2I_d)$, Oracle Access to $SDM$, bounds ${\sigma}_{lo}$and ${\sigma}_{hi}$, a number of steps $stps$, a number of samples $sam$ and max collision rate $cr$
\ENSURE Fine-tuned ${\sigma}_{opt}$ for circumventing the $SDM$ with a collision rate of $cr$

\FOR{$stps$ steps of binary search}
    \STATE ${\sigma}_{mid}$ $\leftarrow$ (${\sigma}_{lo}$ + ${\sigma}_{hi}$) / 2
    \STATE Generate $sam$ samples from $\mathcal{N}(0, \sigma^2_{mid}I_d)$
    \STATE Query the SDM $sam$ times
    \STATE collision rate $\leftarrow$ ratio of rejected queries
    \IF{collision rate $ > cr$} 
        \STATE Select the upper half range, $\sigma_{lo} \leftarrow \sigma_{mid}$
    \ELSE
        \STATE Select the lower half range, $\sigma_{hi} \leftarrow \sigma_{mid}$
    \ENDIF
\ENDFOR
\STATE $\sigma_{opt} \leftarrow \sigma_{hi}$ 
\STATE \textbf{return} $\sigma_{opt}$
\end{algorithmic}\label{alg:adapt_proposal_grad_est}
\end{algorithm}
\begin{algorithm}[t]
\caption{\name: Oracle-guided Adaptive Rejection Sampling for the gradient estimation attack element.}
\begin{algorithmic}[1]
\REQUIRE Current sample $x$, Proposal Distribution $\mathcal{N}(0, \sigma^2I_d)$ around current sample, Oracle Access of $SDM$, number of samples $n$, max retries $ge_{tries}$
\ENSURE Model output for $n$ sampled queries

\STATE $ $
Optimal Parameter $\sigma_{opt} \leftarrow$ \texttt{ADAPT\_PROPOSAL}$(\mathcal{N}(0,\sigma^2I_d))$

\WHILE{\# successful queries < $n$ AND \# queries < $ge_{tries}$}
    \STATE Resample $ q \leftarrow \mathcal{N}(0, \sigma_{opt}^2I_d)$
    \STATE Query the model using $x + q$ 
\ENDWHILE
\STATE \textbf{return} successfully sampled queries
\end{algorithmic}\label{alg:reject_sample_grad_est}
\end{algorithm}

\name thus includes a two-pronged \textit{adapt and resample} strategy. First, the adversary \textit{adapts} the proposal distribution, and samples from this distribution to construct attack queries. If, at any stage, a query causes a collision, the adversary can simply \textit{resample} a substitute query from the proposal distribution. \name is a general approach that applies to any attack consisting of the above three attack elements. It does not make any assumptions about the SDM being deployed --- the SDM remains in place at all times and \name must handle any rejections/bans. In the rest of this section, we describe our strategy for choosing the proposal distributions. Then, we describe how three standard attack components can be adapted using \name to attack SDMs successfully.

\subsubsection{\textbf{Strategy for Choosing Proposal Distributions.}}
\label{sec:prop_dis}
Our rationale for choosing the proposal distributions is to minimize changes to the vanilla attacks as much as possible. We note that in vanilla versions of the attacks, each attack element either samples (1) from a fixed distribution (e.g., Gaussian with fixed variance) or (2) based on a fixed parameter (e.g., fixed step-size).  As such, two cases arise:

\begin{itemize}
    \item The first case is when performing stochastic operations, such as sampling from a Gaussian distribution or a uniform distribution. In such a case, \name adopts the same distribution with an important change: adapt the parameters of that distribution. For example, since base/vanilla NES samples from a Gaussian distribution for gradient estimation, \name-NES assumes a Gaussian with tunable variance as the proposal distribution, adapts the variance, and then samples from it. Another example is the HSJA attack which uniformly samples from a sphere of a fixed radius. \name-HSJA uses a uniform distribution of the sphere radius as the proposal distribution.
    
    \item The second case is when queries are based on a fixed parameter. Here, we construct the proposal distribution as uniformly distributed in the subspace of the queries resulting from the operation with the appropriate tunable parameter. For example, when taking a step, vanilla NES takes a fixed-sized step in the direction of the sign of the gradient. The proposal distribution in \name-NES samples queries uniformly over all possible step directions, captured by the Rademacher distribution parameterized by step size. After that, the samples are scaled by the step size (the tunable parameter which is adapted during the adapt phase).
\end{itemize}

\subsubsection{\textbf{Estimating a gradient.}}\label{sec:est_grad}
 Gradient estimation typically samples points around the current sample ($x_t$) by adding a noise term from a distribution such as an isotropic Gaussian $\mathcal{N}(0,\sigma^2 I_d)$ (parameterized by $\sigma$) or uniform distribution on a sphere. A small value of $\sigma$ results in sampling points within a small neighborhood around $x_t$. In that case, the attack will converge faster at the expense of query collisions. For example, during gradient estimation, NES samples $n$ close-by variants of a given starting point $x_t$. It then performs localized differences to estimate the gradient.

\paragraph{\textbf{\name Variant.}}
We apply the \name-based adapt and resample strategy to construct a proposal distribution for sampling gradient estimation queries that avoid a collision. First, we adapt the sampling distribution used by the underlying attack (e.g., Gaussian with zero mean and parameterized by variance, $\mathcal{N}(0,\sigma^2I_d)$ for NES). However, unlike current attacks that use a fixed $\sigma$, we \textit{adapt} the proposal distribution by performing an oracle-guided fine-tuning of $\sigma$ using binary search. The entire process can be seen in Algorithm~\ref{alg:adapt_proposal_grad_est} and illustrated in Figure~\ref{fig:hsja_ge_adapt} for NES. Specifically, the binary search samples $sam$ queries at each point evaluated between a given $\sigma_{lo}$ and $\sigma_{hi}$. The search is run for $stps$ steps, converging at the smallest distance that results in less than the tolerable collision rate $cr$. Finally, one can sample the gradient estimation samples from $\mathcal{N}(0,\sigma_{opt}^2I_d)$ where ${\sigma_{opt}}^2$ is the fine-tuned variance. A similar process can be used if a different distribution is used instead of a Gaussian.

Second, we perform rejection sampling using the fine-tuned proposal distribution to get the model output for $n$ points, which are then used to estimate the gradient by averaging the loss differences at the $n$ samples. To deal with possible collisions, our strategy is to \emph{resample} from the proposal distribution up to a given $ge_{tries}$ times to ensure we get all $n$ valid samples. Algorithm~\ref{alg:reject_sample_grad_est} details this process (and Figure~\ref{fig:hsja_ge_using_proposal} also illustrates it for NES). However, even after trying for $ge_{tries}$ times, if we still do not have $n$ valid samples, we move forward - as long as we have 1 valid sample, we average over the valid samples we have. Figure~\ref{fig:hsja_ge_adapt} illustrates the adaptive gradient estimation for NES.

\subsubsection{\textbf{Taking a Step.}}\label{sec:taking_a_step}{
The step size for this attack element presents a trade-off between collision and convergence. For example, HSJA takes the largest step size possible (such that the resultant point is still adversarial) and decreases this value over time to speed up convergence. However, if the attack queries use a small step, the adversary risks a collision and could fail to make progress toward generating an adversarial example.}

\paragraph{ \textbf{\name Variant.}} 
We apply the \name adapt and resample strategy to construct a proposal distribution to sample points using the smallest step that avoids collisions. 
First, we assume a proposal distribution that models steps taken by the attack. For instance, a suitable distribution could be (a) a $\lambda-$scaled Rademacher distribution for the NES attack, to represent steps taken towards the sign of the gradient or (b) a uniform distribution on the surface of a sphere of radius $\lambda$ for the HSJA attack, to represent steps taken in the direction of the gradient. Here, $\lambda$ is the tunable step size.
Then, much like Algorithm~\ref{alg:adapt_proposal_grad_est} adapts a Gaussian for gradient estimation, we \textit{adapt} the proposal distribution by performing an Oracle-guided fine-tuning of $\lambda$ using binary search resulting in $\lambda_{opt}$. Finally, given a gradient direction, one can sample the step from the proposal distribution conditioned on the gradient direction.
 Second, in the case of a collision, we \textit{resample} the corresponding step up to $steps_{tries}$ times along different gradient directions (similar to the resampling procedure for gradient estimation in Algorithm~\ref{alg:reject_sample_grad_est}).  

\subsubsection{\textbf{Locating the Boundary.}}\label{sec:boundary_location}{Recall that the attacker locates a decision boundary by interpolating along a line from the current attack sample towards the original input sample. This interpolation is a binary search where the minimum distance between successive queries decreases exponentially. If this distance is too small at any point during this interpolation, the query may collide and prevent attack progress. 

\paragraph{\textbf{\name Variant.}}{
 We define a binary search operation: $\phi(r)$, which proceeds until the distance between successive queries is less than the distance $r$, which we define as the termination distance. We apply the \name adapt and resample strategy to construct a proposal distribution for sampling the termination distance for the interpolation binary search. First, we use a uniform proposal distribution parameterized by the upper bound $a$ : $U(0, a)$. This upper bound is initialized by the distance between current sample $x_t$ and victim sample $x_{vic}$. Then, we \textit{adapt} the proposal distribution by performing an oracle-guided fine-tuning of $a$ using binary search. Second, in case of an unsuccessful binary search, we resample another termination distance and perform the binary search $\phi(r)$ again. In the implementation, we efficiently perform this \textit{resampling} by reusing the unsuccessful binary search and taking the penultimate query (equivalent to a binary search operation with twice the termination distance).}

\subsubsection{\textbf{Determining SDM Store}}
\label{sec:att_extract}
We observe that the cost of including our \name strategy depends on the specific SDM being deployed. For query rejection-based defenses, i.e., Blacklight~\cite{li2022blacklight}, PIHA~\cite{choi2023piha} and IIoT-SDA~\cite{esmaeili2022iiot}, the additional query overhead from \name can be run without incurring additional account requirements. However, for account banning-based defenses like OSD~\cite{chen2020stateful}, the additional overhead may require more accounts than is needed to perform the standard attack across multiple steps (Section~\ref{sec:eval_vanilla}). As such, we devise a test to determine 1) if the defense is banning accounts and 2) if it is using a per-account or global query store. This test has two steps. First, the attack creates an account A and feeds it in an input $x$. It queries  $x$ until it is detected on account A (or up until a certain limit), suggesting that SDMs are likely not being used. Second, the attacker creates account B and queries $x$ again on that account (e.g., a query that was detected using account A). If it is detected, the system must be using a global query store. Otherwise, we assume that it is using a per-account scheme. The cost of this test is the number of queries required for detection on A, the additional query on B, and the additional account. 

\subsection{Modifying Existing Attacks with \name}\label{sec:att_existing}
We now describe how to modify six existing black-box attacks to demonstrate how to apply our adaptations to a variety of attacks. Figure~\ref{fig:examples} shows diagrams visualizing the changes for the gradient estimation stage of NES~\cite{ilyas2018black}. Here, we describe the attack components that are modified when integrated with \name. We also summarize which elements are modified for the specific attacks we analyze in Table~\ref{tab:att_sum}. Source code for our modified black-box attacks is available at \MYhref[magenta]{https://github.com/nmangaokar/ccs\_23_oars\_stateful\_attacks}{https://github.com/nmangaokar/ccs\_23\_oars\_stateful\_attacks}.

While we demonstrate specific modifications for six attacks, we note that the \name approach of adapting proposal distributions and resampling queries generalizes to attacks with the three elements discussed in Section~\ref{sec:bb_elements}. The discussions below cover a range of differing attacks and provide blueprints for adapting other attacks. This is done by first identifying the three common elements and applying the corresponding modifications to adapt proposal distributions and resample as applicable.

\begin{table}[t]
\centering
\caption{Table summarizing the adaptive modifications to these attacks. Full circle means that we apply \name's adapt and resample to modify this element. Empty circle means that the attack does not have this element. }\label{tab:att_sum}
\begin{tabular}{lccc}
\toprule
\textbf{Attack}  & \multicolumn{1}{c}{\textbf{\begin{tabular}[c]{@{}c@{}}Est. a \\ Gradient\end{tabular}}} & \multicolumn{1}{c}{\textbf{\begin{tabular}[c]{@{}c@{}}Taking a \\ Step\end{tabular}}} & \multicolumn{1}{c}{\textbf{\begin{tabular}[c]{@{}c@{}}Locating the \\ Boundary\end{tabular}}} \\ \midrule
\textbf{\name-NES}      &      \includegraphics[width=0.012\textwidth]{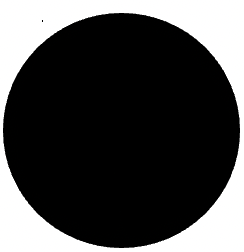}                                         &      \includegraphics[width=0.012\textwidth]{figs/full_circle_hack.png}                                      &         \includegraphics[width=0.012\textwidth]{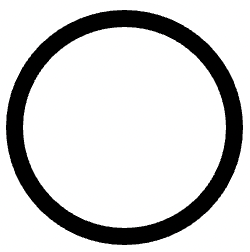}                                         \\
\textbf{\name-Square }   &   \includegraphics[width=0.012\textwidth]{figs/empty_circle_hack.png}                                                  &   \includegraphics[width=0.012\textwidth]{figs/full_circle_hack.png}                                          &      \includegraphics[width=0.012\textwidth]{figs/empty_circle_hack.png}                                                \\
\textbf{\name-HSJA }     &          \includegraphics[width=0.012\textwidth]{figs/full_circle_hack.png}                                           &     \includegraphics[width=0.012\textwidth]{figs/full_circle_hack.png}                                         &    \includegraphics[width=0.012\textwidth]{figs/full_circle_hack.png}                                                  \\
\textbf{\name-QEBA }     &                                                   \includegraphics[width=0.012\textwidth]{figs/full_circle_hack.png}                                           &     \includegraphics[width=0.012\textwidth]{figs/full_circle_hack.png}                                         &    \includegraphics[width=0.012\textwidth]{figs/full_circle_hack.png}                                                    \\
\textbf{\name-SurFree }  &      \includegraphics[width=0.012\textwidth]{figs/empty_circle_hack.png}                                               &    \includegraphics[width=0.012\textwidth]{figs/full_circle_hack.png}                                          &     \includegraphics[width=0.012\textwidth]{figs/full_circle_hack.png}                                                \\
\textbf{\name-Boundary} &     \includegraphics[width=0.012\textwidth]{figs/empty_circle_hack.png}                                                 &   \includegraphics[width=0.012\textwidth]{figs/full_circle_hack.png}                                            &      \includegraphics[width=0.012\textwidth]{figs/full_circle_hack.png}                                                 \\ \bottomrule
\end{tabular}

\end{table}

\subsubsection{\textbf{\name-NES (Score-based).}}
To create \name-NES, we use \name to modify vanilla NES's~\cite{ilyas2018black} gradient estimation (estimating a gradient) and projected gradient descent (taking a step) to avoid detection with more dissimilar queries.

\paragraph{\textbf{Estimating a Gradient.}} {

Vanilla NES~\cite{ilyas2018black} estimates a gradient using finite differences over $n$ Monte Carlo samples from a Gaussian distribution as shown below, where $L$ is the loss function to be estimated, $\sigma^2$ is the variance of the distribution, and $u_i \sim \mathcal{N}(0, I)$: 
\begin{equation}\label{eqn:nes_smooth_loss_grad}
    \nabla^{est}_{x_t} L =\frac{1}{\sigma n}\sum_{i=1}^n\left[ L(x_t + \sigma u_i,~y) u_i\right]
\end{equation}

\underline{Modified}: \textit{We use a Gaussian proposal distribution $\mathcal{N}(0,\sigma^2I_d)$ to sample the gradient estimation queries that avoid collision. Then, we use \name's adapt \& resample to fine-tune $\sigma$ and generate $n$ queries for gradient estimation. This is similar to the approach described in Section \ref{sec:est_grad}}
}

\paragraph{\textbf{Taking a Step.}}{

For taking a step, vanilla NES updates with a fixed step size $\lambda$ with projected gradient descent: 
\begin{equation}\label{eqn:nes_pgd}
    x_{t+1} = \text{Proj}_{x_{orig} + \mathcal{S}}(x_t + \lambda \cdot \text{sgn}(\nabla^{est}_{x_t} L))
\end{equation}
where $\mathcal{S} = \{\delta: ||\delta|| \leq \epsilon\}$ and Proj is the projection operator.

\underline{Modified}: \textit{We use a $\lambda$-scaled Rademacher distribution where $0 \leq \lambda \leq \epsilon$ to sample the smallest step queries that avoid collision. Then, we use \name's adapt \& resample to fine-tune $\lambda$ and sample the next step. This is similar to the approach described in Section \ref{sec:taking_a_step}}
}

\subsubsection{\textbf{\name-Square (Score-based).}}
To create \name-Square, we use \name to modify the number of the random square perturbations being sampled at each step. Square attack does not have a gradient estimation step.

\paragraph{\textbf{Taking a Step.}} {

The $\ell_\infty$ version of vanilla Square~\cite{andriushchenko2020square} samples random squares with perturbations filled to $-\epsilon$ or $\epsilon$ and selecting the square that increases the loss function. In particular, it samples random squares of progressively smaller sizes to help the attack converge.

\underline{Modified}: \textit{We apply \name's adapt and resample for sampling squares while evading detection by the SDM. For each square size, we first modify the sampling process of the Square attack to select multiple squares for each step. Then, we construct a proposal distribution for sampling the minimum number of squares required to avoid a collision. We then apply \name's adapt mechanism to fine-tune the proposal distribution. Finally, we use \name's resample mechanism to handle collisions. When the attack proceeds to a smaller square size, we repeat the process.}
}

\subsubsection{\textbf{\name-HSJA (Hard-label).}}
To create \name-HSJA, we use \name to modify vanilla HSJA's~\cite{chen2020hopskipjumpattack} gradient estimation (estimating a gradient), geometric progression of step size (taking a step), and boundary search (locating the boundary) operations to avoid collisions. 

\paragraph{\textbf{Estimating a Gradient}} {
Gradient estimation works similarly to NES~\cite{ilyas2018black} but with a variance-reduced estimate over a different loss function as described below per the equations in the original paper:
\begin{equation}
    \nabla_x^{est} L = \frac{1}{\zeta n}\sum_{i=1}^n (\phi_{x_{vic}}(x_t + \zeta u_i) - \overline{\phi_{x_{vic}}})u_i,
\end{equation}
where
\begin{equation}
    \overline{\phi_{x_{vic}}} = \frac{1}{n}\sum_{i=1}^n \phi_{x_{vic}}(x_t + \zeta u_i),
\end{equation} $\phi_{x_{vic}}(x)$ is 1 if $x$ is adversarial with respect to $x_{vic}$ and -1 if it is not, $u_i$ is a randomly sampled vector from the uniform distribution, and $\zeta$ is a small positive hyperparameter.

\underline{Modified}: \textit{To match the underlying attack distribution, we use a proposal distribution described by the uniform distribution on the surface of a sphere with radius $\zeta$ to sample the gradient estimation queries that avoid a collision. Then, we use \name's adapt \& resample to fine-tune $\zeta$ and generate $n$ queries for gradient estimation.}
}

\paragraph{\textbf{Taking a Step}}{

For taking a step, vanilla HSJA~\cite{chen2020hopskipjumpattack} sets the initial step size to:
\begin{equation}
    \lambda = \|x_t - x_{vic}\|_p/\sqrt{t}
\end{equation}

where $p$ represents the targeted threat model $\ell_p$ norm. HSJA then applies geometric progression, halving the step size until an adversarial example is found. 

\underline{Modified}: \textit{We use a proposal distribution described by the uniform distribution on the surface of a sphere with radius $\lambda$ to sample the smallest step queries that avoid a collision. Then, we use \name's adapt \& resample to sample the next step.}
}

\paragraph{\textbf{Locating the Boundary.}} {
Vanilla HSJA~\cite{chen2020hopskipjumpattack} applies binary search between the post-step point and the original victim point to locate the boundary.

\underline{Modified}: \textit{Similar to the approach described in \ref{sec:boundary_location}, we define a binary search with a termination distance and use a uniform proposal distribution parameterized by the upper bound. Then we use \name's adapt \& resample to proceed with the attack.}
}

\subsubsection{\textbf{\name-QEBA (Hard-label).}}
The changes for \name-QEBA are the same as that of \name-HSJA - the only algorithmic difference between vanilla QEBA~\cite{li2020qeba} and vanilla HSJA~\cite{chen2020hopskipjumpattack} is using a lower dimensionality to sample from, and this algorithmic change does not necessitate an additional modification in how we apply \name to HSJA. 

\subsubsection{\textbf{\name-SurFree (Hard-label).}}
To create \name-SurFree, we use \name to modify two elements of vanilla SurFree~\cite{maho2021surfree} : magnitude adjustment of the polar coordinate representation (taking a step), and the final boundary location refinement between the proposed attack and the original victim sample (locating the boundary) to avoid collisions.

%To create our Adaptive SurFree attack, we use \name to modify two elements of vanilla SurFree~\cite{maho2021surfree} : magnitude adjustment of the polar coordinate representation (taking a step), and the final boundary location refinement between the proposed attack and the original victim sample (locating the boundary) to avoid collisions.

\paragraph{\textbf{Taking a Step.}} {

Vanilla SurFree~\cite{maho2021surfree} relies on random search over directions parameterized by polar coordinates. In particular, given the original sample $x_{vic}$, the current attack iteration example $x_t$ (with $u$ defined as $x_{vic} - x_t$), SurFree tries to find new directions such that the boundary is closer. First, it uses Gram-Schmidt to sample a random vector $v$ that is orthogonal to $u$. Then, SurFree computes a weighted addition of $u$ and $v$ described by the polar coordinate $\alpha$. 

\underline{Modified}: \textit{We use a proposal distribution described by the uniform distribution on the surface of a sphere with radius $\lambda$ to sample the smallest step in the polar coordinate that avoids collision. Then, we use \name's adapt \& resample to sample the next step.}
}

\paragraph{\textbf{Locating the Boundary.}}
{

For locating the boundary, vanilla SurFree~\cite{maho2021surfree} applies a binary search between the final $x_t$ and $x_{vic}$. Like \name-HSJA, we apply \name as described in Section~\ref{sec:boundary_location}.
}

\subsubsection{\textbf{\name-Boundary (Hard-label).}}

In order to create \name-Boundary, we use \name to modify vanilla Boundary's~\cite{brendel2017decision} to adapt and resample to evade collisions. Our application of \name adjusts Boundary's adjustment of the hyperparameters $\eta_\delta$ (to control the distance to move in the random direction being searched - i.e., taking a step) and $\eta_\epsilon$ (to control the distance being moved back to the original point - i.e., locating the boundary).

\paragraph{\textbf{Taking a Step.}}{

As vanilla Boundary~\cite{brendel2017decision} performs its random boundary walk, it iteratively operates two stages, with the first stage exploring $k$ random orthogonal directions scaled by a factor of $0 < \eta_\delta < 1$, starting from an adversarial point. Boundary updates the value of $\eta_\delta$ using Trust Region methods: Compute the ratio of adversarial samples to the total samples; Increase $\eta_\delta$ if the ratio is too high; else, decrease it. 

\underline{Modified:} \textit{We use a proposal distribution described by the uniform distribution on the surface of a sphere with radius $\lambda$ to sample the smallest scaling factor that avoids collision. Then, we use \name's adapt \& resample to sample the next step.}
}

\paragraph{\textbf{Locating the Boundary.}} {

In the second stage of vanilla Boundary~\cite{brendel2017decision}, the algorithm then takes the adversarial points from the first stage and move back up to boundary towards the original victim sample $\eta_\epsilon$. The value of $\eta_\epsilon$ is again adjusted with Trust Region methods, increasing $\epsilon$ if too many samples are adversarial and decreasing $\eta_\epsilon$ if there are too few adversarial samples to match a given threshold. We apply \name to this stage by using the approach described in Section \ref{sec:boundary_location}.
}

\section{Experiments}\label{sec:experiments}
We evaluate our six \name black-box attacks against four state-of-the-art SDMs: Blacklight~\cite{li2022blacklight}, PIHA~\cite{choi2023piha}, IIoT-SDA~\cite{esmaeili2022iiot}, and OSD~\cite{chen2020stateful}. Below, we list three questions that our evaluation answers, along with a summary of the key takeaways: \smallskip

\noindent
$\bullet$ \textbf{Are \name attacks successful against vanilla SDMs?}\\
We find our set of \name attacks to be more successful than the existing query-blinding and standard attack baselines on the CIFAR10~\cite{krizhevsky2009learning}, ImageNet~\cite{russakovsky2015imagenet}, CelebaHQ~\cite{karras2018progressive}, and IIoT Malware~\cite{esmaeili2022iiot} datasets, which each defense suffering from at least one attack that achieves a 99\% attack success rate. (Section~\ref{sec:eval_vanilla})\smallskip

\noindent
$\bullet$ \textbf{Do \name attacks suffer when reconfiguring SDMs?}\\
We consider different approaches SDMs might consider to thwart these attacks (e.g., different defense hyperparameters). We show that our attacks continue to adapt to these changes and successfully construct adversarial examples. (Section~\ref{sec:reconfigs_sdm}) \smallskip

\noindent
$\bullet$ \textbf{What is the incurred attacker cost in the presence of SDM?}\\
We report the number of needed queries for successful \name attacks under different configurations. This includes queries incurred during both stages of \name. Our findings indicate that these SDMs adaptations increase the number of queries required by the attacker. However, the cost of added queries is only $\$$1-18 for online APIs. (Section~\ref{sec:eval_cost})

\subsection{Experimental Setup}
\label{sec:exp-setup}
\paragraph{\textbf{Defenses and Classification Tasks}.}
We test Blacklight~\cite{li2022blacklight} and PIHA~\cite{choi2023piha} on three image classification tasks: CIFAR10~\cite{krizhevsky2009learning} and ImageNet~\cite{russakovsky2015imagenet} for object classification and CelebaHQ~\cite{karras2018progressive} for identity classification. We used ResNet~\cite{he2016deep} based models for these three datasets. We evaluate OSD~\cite{chen2020stateful} only on CIFAR10 as that is the only dataset it is available for. Finally, we evaluate IIoT-SDA~\cite{esmaeili2022iiot} on the IIoT malware classification task~\cite{azmoodeh2018robust} used in the original paper. Table~\ref{tab:datasets_models} contains additional information about the datasets and the corresponding classification models. 

\begin{table}
\caption{Overview of classification tasks, their datasets, classifiers, and test set accuracy.}
\label{tab:datasets_models}
\begin{tabular}{ccccc}
\toprule
\textbf{Dataset} & \textbf{Classes} & \textbf{Model} & \textbf{Inp. Shape} & \textbf{Acc.} \\ \midrule
CIFAR10 & 10 & ResNet-20 & 32x32x3 & 91.73\% \\
ImageNet & 1000 & ResNet-152 & 224x224x3 & 78.31\% \\
CelebaHQ & 307 & ResNet-152 & 256x256x3 & 89.55\% \\
IIoT & 2 & 5 Conv + FC & 224x224x1  & 97.46\% \\ \bottomrule
\end{tabular}

\end{table}

\paragraph{\textbf{Defense Configurations.}} In Section~\ref{sec:eval_vanilla}, we evaluate the vanilla SDMs under their default and originally proposed configurations. Specifically, Blacklight~\cite{li2022blacklight} utilizes a window size of 20 (for CIFAR10~\cite{krizhevsky2009learning}, or 50 for ImageNet~\cite{russakovsky2015imagenet} and CelebaHQ~\cite{karras2018progressive}), a quantization step of 50, and a threshold of 0.5. PIHA~\cite{choi2023piha} uses a block size of 7x7, and a threshold of 0.05. OSD~\cite{chen2020stateful} uses 50 nearest neighbors and a threshold of 1.44. IIoT~\cite{esmaeili2022iiot} uses 11 nearest neighbors and a threshold of 0.21. We change these configurations in section~\ref{sec:reconfigs_sdm} to evaluate the attacks in under modified versions of these SDMs.

\paragraph{\textbf{Attack Configurations}.}
We attack the above defenses with variants of the NES~\cite{ilyas2018black}, Square~\cite{andriushchenko2020square}, HSJA~\cite{chen2020hopskipjumpattack}, QEBA~\cite{li2020qeba}, SurFree~\cite{maho2021surfree}, and Boundary~\cite{brendel2017decision} attacks described in Sections~\ref{sec:bb_attacks}. For each attack's vanilla hyperparameters, we employ the default values. NES and Square are $\ell_\infty$ score-based attacks, and HSJA, QEBA, SurFree, and Boundary are $\ell_2$ hard-label attacks. Furthermore, NES, HSJA, and QEBA are targeted attacks; the remaining attacks are untargeted. For CIFAR10, ImageNet, and CelebaHQ, we use the standard $\ell_\infty$ and normalized $\ell_2$ $\epsilon=0.05$ perturbation budgets used in prior work~\cite{li2022blacklight}. For IIoT malware, we use the $\epsilon=0.2$ budget employed by prior work~\cite{esmaeili2022iiot}. Like in prior work~\cite{li2022blacklight}, we also assume a query budget of $100$k.

We run all attacks under two baseline configurations and three of our attack configurations. The first baseline is the standard configuration, which runs the attacks until success, query budget exhaustion, or a collision occurs. Note that this is the setting evaluated by prior work~\cite{li2022blacklight}. The second baseline is the query-blinding configuration, which builds upon the standard configuration by applying query-blinding~\cite{chen2020stateful}. We use the standard random affine transformation from prior work~\cite{li2022blacklight} for query-blinding, where each query is randomly rotated by up to $10^\circ$ degrees, shifted by up to 10\% horizontally/vertically, and zoomed by up to 10\%. Again, attacks are run until success, budget exhaustion, or rejection. These transformations also apply to the data from the IIoT Malware dataset, which are represented in an 2d matrix format. 

We evaluate three variants of our \name attack configurations. The first two only apply \name's resample strategy (only rejection sampling without fine-tuning the proposal distribution) to the standard and query blinding configurations. For a given attack, we enable the resample mechanisms described in Section~\ref{sec:approach} for \textit{estimating a gradient}, \textit{taking a step}, and \textit{locating the boundary}. The third of our attack configurations includes the full recommended \name configuration that enables both adapt and resample mechanisms (rejection sampling with proposal fine-tuning). Table~\ref{tab:adaptive_hyperparams} in Appendix~\ref{sec:app_hyper} includes the detailed hyperparameters for all three configurations. We evaluate all five attack configurations against 100 samples, with target classes chosen uniformly at random for targeted attacks. For CIFAR10, we additionally report results on 1000 samples, included in Table~\ref{tab:cifar1000} in Appendix~\ref{sec:app_add_eval}.

\begin{table*}[t]
\small
\centering
 \caption{Our proposed adaptive attacks with adapt and resample outperform existing standard and query-blinding baselines. Results are presented in (ASR / query count) format. We find that with our adapt and resample techniques each dataset and defense combination suffers from at least one attack that achieves 99\% or higher ASR. We also show that some attacks can show some improvement with just resampling mechanisms. Bold numbers are the best hard-label attacks, and bold italicized numbers are the best score-based attacks.}\label{tab:exp_global}

\begin{tabular}{lllcccccc}
\toprule
\multirow{2}{*}{\textbf{\begin{tabular}[c]{@{}c@{}}\\Dataset\end{tabular}}} & \multirow{2}{*}{\textbf{\begin{tabular}[c]{@{}c@{}}\\Defense\end{tabular}}} & \multirow{2}{*}{\textbf{\begin{tabular}[c]{@{}c@{}} \\ Attack\end{tabular}}} & \multirow{2}{*}{\textbf{\begin{tabular}[c]{@{}c@{}} \\ Targeted\end{tabular}}} & \multicolumn{2}{c}{\textbf{Baseline}} & \multicolumn{3}{c}{\textbf{Proposed}} \\\cmidrule(lr){5-6}\cmidrule(l){7-9}
& & & & \textbf{Standard} & \textbf{\begin{tabular}[c]{@{}c@{}}Query\\ Blinding\end{tabular}} & \textbf{\begin{tabular}[c]{@{}c@{}}Standard\\ + Resample\end{tabular}} & \textbf{\begin{tabular}[c]{@{}c@{}}Query Blinding\\ + Resample\end{tabular}}  & \textbf{\begin{tabular}[c]{@{}c@{}}Adapt\\ + Resample\end{tabular}} \\ 
 &  &  &  & 4& 5& 6& 7&8\\ \midrule
\multicolumn{1}{l|}{\multirow{12}{*}{\textbf{CIFAR10}}}  & \multirow{6}{*}{\textbf{Blacklight}} & \textbf{NES}  & \checkmark                                                 & 0\% / - &   0\% / -                                  &       0\% / -                                                    &             4\% / 959                                             &       \textbf{\textit{99\% / 1540}}        \\
\multicolumn{1}{l|}{}                                    &                                      & \textbf{Square}  &                                              &         0\% / -          &             33\% / 2                                              &         0\% / -                                                  &  34\% / 11                                                         &                                                                      93\% / 218        \\
\multicolumn{1}{l|}{}                                    &                                      & \textbf{HSJA} & \checkmark                                                 &         0\% / -          &                    0\% / -                                       &            0\% / -                                               &      0\% / -                                                     &                                                                       82\% / 1615       \\
\multicolumn{1}{l|}{}                                    &                                      & \textbf{QEBA}  & \checkmark                                                 &         0\% / -          &                0\% / -                                           &             0\% / -                                              &      0\% / -                                                     &                                                                    \bf{98\% / 1294}          \\
\multicolumn{1}{l|}{}                                    &                                      & \textbf{SurFree}    &                                            &         0\% / -          &                    1\% / 19                                       &               79\% / 48                                            &      4\% / 21                                                     &                                                                      81\% / 145        \\
\multicolumn{1}{l|}{}                                    &                                      & \textbf{Boundary}     &                                         &           0\% / -        &                                   0\% / -                       &                           7\% / 682       & 6\% / 1449 & \textbf{98\% / 3302}                                                                                                       \\ \cmidrule(l){2-9} 
\multicolumn{1}{l|}{}                                    & \multirow{6}{*}{\textbf{PIHA}}       & \textbf{NES}      & \checkmark                                             &         0\% / -          &    0\% / -                                                       &   2\% / 374                                                        &          5\% / 369                                                 &                                                                    83\% / 1646        \\
\multicolumn{1}{l|}{}                                    &                                      & \textbf{Square}    &                                             &        29\% / 3           &               35\% / 2                                            &                      90\% / 101                                     &     38\% / 3                                                      &                                                                       \textbf{\textit{99\% / 191}}       \\
\multicolumn{1}{l|}{}                                    &                                      & \textbf{HSJA}     & \checkmark                                             &         0\% / -          &     0\% / -                                                      &       0\% / -                                                    &      0\% / -                                                    &                                                                     76\% / 2811        \\
\multicolumn{1}{l|}{}                                    &                                      & \textbf{QEBA}      & \checkmark                                            &        0\% / -           &      0\% / -                                                     &         0\% / -                                                   &        1\% / 14818                                                   &                                                                      \bf{95\% / 1384}        \\
\multicolumn{1}{l|}{}                                    &                                      & \textbf{SurFree}    &                                           &          0\% / -         &    2\% / 24                                                       &     74\% / 44                                                      &      1\% / 19                                                     &                                                                      67\% / 155        \\
\multicolumn{1}{l|}{}                                    &                                      & \textbf{Boundary}     &                                         &         0\% / -          &                                   0\% / -                        &                                      80\% / 719                     &                                                 74\% / 729          &                    {90\% / 915 }                                                         \\ \midrule
\multicolumn{1}{l|}{\multirow{12}{*}{\textbf{ImageNet}}} & \multirow{6}{*}{\textbf{Blacklight}} & \textbf{NES}       & \checkmark                                            &            0\% / -       &         0\% / -                                                  &          0\% / -                                                &     0\% / -                                                      &                                                                     \textbf{\textit{100\% / 10551}}        \\
\multicolumn{1}{l|}{}                                    &                                      & \textbf{Square}    &                                            &         0\% / -          &               25\% / 2                                             &         0\% / -                                                  &              25\% / 2                                             &                                                                       84\% / 173       \\
\multicolumn{1}{l|}{}                                    &                                      & \textbf{HSJA}           & \checkmark                                       &       0\% / -            &      0\% / -                                                     &                0\% / -                                           &                                   0\% / -                        &                                                                   50\% / 27620           \\
\multicolumn{1}{l|}{}                                    &                                      & \textbf{QEBA}    & \checkmark                                              &          0\% / -         &       0\% / -                                                    &              0\% / -                                             &                                    0\% / -                       &                                                                  50\% / 37924           \\
\multicolumn{1}{l|}{}                                    &                                      & \textbf{SurFree}      &                                         &         0\% / -          &                  0\% / -                                         &                  71\% / 204                                        &                   0\% / -                                        &                                                                     \textbf{93\% / 1006}       \\
\multicolumn{1}{l|}{}                                    &                                      & \textbf{Boundary}   &                                           &            0\% / -       &                                                    0\% / -       &                                                  28\% / 1024         &                                           24\% / 759                &               38\% / 4262                                                               \\ \cmidrule(l){2-9} 
\multicolumn{1}{l|}{}                                    & \multirow{6}{*}{\textbf{PIHA}}       & \textbf{NES}       & \checkmark                                            &        14\% / 9283           &      0\% / -                                                     &        34\% / 8010                                                  &       0\% / -                                                    &                                                                       \textbf{\textit{92\% / 8578}}      \\
\multicolumn{1}{l|}{}                                    &                                      & \textbf{Square}  &                                              &       28\% / 3            &                 22\% /   2                                        &      34\% / 4                                                     &       21\% / 2                                                   &                                                                    87\% / 203         \\
\multicolumn{1}{l|}{}                                    &                                      & \textbf{HSJA}     & \checkmark                                             &        0\% / -           &      0\% / -                                                     &        0\% / -                                                  &                                                                                         0\% / -                          &    91\% / 29998           \\
\multicolumn{1}{l|}{}                                    &                                      & \textbf{QEBA}     & \checkmark                                             &        0\% / -           &       0\% / -                                                    &           0\% / -                                              &                                                                                    0\% / -                                 &     96\% / 22947          \\
\multicolumn{1}{l|}{}                                    &                                      & \textbf{SurFree}  &                                             &     0\% / -              & 0\% / -                                                          &         \textbf{100\% / 555}                                                 &              0\% / -                                              &                                                                  \textbf{100\% / 1892}           \\
\multicolumn{1}{l|}{}                                    &                                      & \textbf{Boundary}   &                                           &           0\% / -        &                                       0\% / -                    &                  35\% / 1980                                         &                                 32\% / 1994                          &                                          46\% / 1278                                    \\ \midrule
\multicolumn{1}{l|}{\multirow{12}{*}{\textbf{CelebaHQ}}}   & \multirow{6}{*}{\textbf{Blacklight}} & \textbf{NES}   & \checkmark                                                &        0\% / -            &          0\% / -                                                  &          0\% / -                                                  &                 0\% / -                                          &                                                                            \textbf{\textit{100\% / 7719}}  \\
\multicolumn{1}{l|}{}                                    &                                      & \textbf{Square}   &                                             &  0\% / -                 & 14\% / 2                                                          &           0\% / -                                                 &             14\% / 2                                             &                                                                  96\% / 211           \\
\multicolumn{1}{l|}{}                                    &                                      & \textbf{HSJA}    & \checkmark                                              &   0\% / -                &    0\% / -                                                        &       0\% / -                                                   &           0\% / -                                                &                                                                     77\% / 31512          \\
\multicolumn{1}{l|}{}                                    &                                      & \textbf{QEBA}   & \checkmark                                               &       0\% / -            &           0\% / -                                                &        0\% / -                                                   &         0\% / -                                                  &         93\% / 8931                                                                   \\
\multicolumn{1}{l|}{}                                    &                                      & \textbf{SurFree}   &                                            &     0\% / -              &   0\% / -                                                        &           93\% / 56                                                &               0\% / -                                            &                                                                      \textbf{98\% / 167}        \\
\multicolumn{1}{l|}{}                                    &                                      & \textbf{Boundary}   &                                           &      0\% / -             &                         0\% / -                                  &             51\% / 697 &      54\% / 1284 &                     73\% /    1974                                                                                                \\ \cmidrule(l){2-9} 
\multicolumn{1}{l|}{}                                    & \multirow{6}{*}{\textbf{PIHA}}       & \textbf{NES}      & \checkmark                                             &    39\% / 7894               &      0\% / -                                                     &        74\% / 7369                                                   &          0\% / -                                                 &                                                                    97\% / 7205         \\
\multicolumn{1}{l|}{}                                    &                                      & \textbf{Square}       &                                         &  23\% / 3                 &     0\% / -                                                      &             33\% / 5                                              &                  14\% / 2                                         &                                                                 \textbf{\textit{100\% / 227}}           \\
\multicolumn{1}{l|}{}                                    &                                      & \textbf{HSJA}     & \checkmark                                             &    0\% / -               &     0\% / -                                                      &        0\%                                                  &                    0\% / -                                      &                                                                  90\% / 30934           \\
\multicolumn{1}{l|}{}                                    &                                      & \textbf{QEBA}   & \checkmark                                               &    0\% / -               &         0\% / -                                                  &         0\% / -                                                  &           0\% / -                                               &      \textbf{100\% / 6984}                                                                        \\
\multicolumn{1}{l|}{}                                    &                                      & \textbf{SurFree}    &                                           &       0\% / -            &     0\% / -                                                      &      \textbf{100\% / 72}                                                     &    0\% / -                                                       &                                                                    \textbf{100\% / 175}          \\
\multicolumn{1}{l|}{}                                    &                                      & \textbf{Boundary}    &                                          &         0\% / -          &                                          0\% / -                 &                          64\% / 684                                 &                               62\% / 876                            &             70\% / 683                                                                 \\ \midrule
\multicolumn{1}{l|}{\multirow{6}{*}{\textbf{\begin{tabular}{@{}c@{}}IIoT \\ Malware\end{tabular}}}}   & \multirow{6}{*}{\textbf{IIoT-SDA}} & \textbf{NES}            & \checkmark                                       &      10\% / 52              &               4\% / 25042                                              &      10\% / 52                                                      &         4\% / 25042                                                  &                                                                     97\% / 3924          \\
\multicolumn{1}{l|}{}                                    &                                      & \textbf{Square}   &                                             &        57\% / 120           &            14\% / 24                                               &         97\% / 221                                                   &          14\% / 24                                                 &                                                                 \textbf{\textit{100\% / 615}}             \\
\multicolumn{1}{l|}{}                                    &                                      & \textbf{HSJA}     & \checkmark                                            &   0\% / -                &            1\% / 80468                                                 &            35\% / 1960                                                &          1\% / 80468                                                  &                                                                      \textbf{100\% / 985}         \\
\multicolumn{1}{l|}{}                                    &                                      & \textbf{QEBA}   & \checkmark                                               &     0\% / -              &         1\% / 48319                                                   &            21\% / 512                                              &             1\% / 48319                                              &        \textbf{100\% / 675}                                                                      \\
\multicolumn{1}{l|}{}                                    &                                      & \textbf{SurFree}     &                                          &     91\% / 210             &                     0\% / -                                       &      91\% / 210                                                     &             0\% / -                                               &          98\% / 455                                                                    \\
\multicolumn{1}{l|}{}                                    &                                      & \textbf{Boundary}    &                                          &         0\% / -          &                                         0\% / -                  &                                     11\% / 395                      &                                     13\% / 414                      &           30\% / 398                                                                \\ \bottomrule
\end{tabular}
\end{table*}

\subsection{Our Adaptive Attacks vs. Existing SDMs}
\label{sec:eval_vanilla}
We now present and compare the attack success rates and query counts of our different attack configurations against the state-of-the-art rejection based defenses, Blacklight~\cite{li2022blacklight}, PIHA~\cite{choi2023piha}, and IIoT-SDA~\cite{esmaeili2022iiot} in
Table~\ref{tab:exp_global}. We report query counts averaged on only successful attacks. We first discuss the results of the baseline attack configurations and then the results of our adaptive \name configurations. We conclude by evaluating OSD~\cite{chen2020stateful}.

\paragraph{\textbf{Baselines}.} Column 4 presents results of the standard configuration, as originally evaluated by the existing SDMs. As expected from prior work~\cite{li2022blacklight}, we find that defenses are consistently able to thwart standard attacks as query collisions almost always occur. Notably, Blacklight~\cite{li2022blacklight} is robust with 0\% ASR for all standard attacks across all datasets. 

The few cases where standard attacks allow for non-zero (but low) ASR are for the PIHA~\cite{choi2023piha} and IIoT-SDA~\cite{esmaeili2022iiot} defenses (e.g., 14\% for NES~\cite{ilyas2018black} against PIHA). The random search-based attack Square~\cite{andriushchenko2020square} produces low but non-zero ASR on all datasets against both PIHA and IIoT-SDA. Since this attack does not rely on gradient estimation and tends to converge very quickly (e.g., 2-120 queries), it has a lower chance of detection as compared to the slower-converging NES~\cite{ilyas2018black}, HSJA~\cite{chen2020hopskipjumpattack}, and QEBA~\cite{li2020qeba} attacks. The gradient estimation procedure of the latter attacks repeatedly queries a large number of nearby points to estimate gradients, adding to its query cost. 

Column 5 presents results of the query blinding~\cite{chen2020stateful} configuration. Again, we find that the query-blinding attacks, as originally evaluated by existing defenses, fail except for a few cases of the Square~\cite{andriushchenko2020square} attack. All other cases of other attacks have $< 5\%$ attack success rate. Manual inspection of the query-blinding attack failures corroborates our discussion in Section~\ref{sec:query-blind}: arbitrary query transformations tend to disrupt the precision of the response, resulting in successful evasion at the cost of aimless ``wandering.'' Overall, we find that neither standard nor query-blinding attacks are sufficient for attacking SDMs.

\paragraph{\textbf{\name}} Columns 6-7 presents results for the \textit{resample-only} configurations (rejection sampling without proposal distribution fine-tuning). Some attacks show moderate improvements. For example, the Square~\cite{andriushchenko2020square} and SurFree~\cite{maho2021surfree} attacks appear to benefit the most from the resample mechanisms (e.g., 100\% for SurFree against PIHA~\cite{choi2023piha} on ImageNet~\cite{russakovsky2015imagenet} and CelebaHQ~\cite{karras2018progressive}). The random search procedures employed instead of gradient estimation are likely to have contributed to this improvement. When combined with the fast convergence properties of a random search, resample mechanisms improve ASR, albeit modestly. However, there is room for more improvement: without adapting to the query collision distribution, the targeted attacks that employ gradient estimation (NES~\cite{ilyas2018black}, HSJA~\cite{chen2020hopskipjumpattack}, QEBA~\cite{li2020qeba}) do not show significant improvement. Once again, adding query-blinding~\cite{chen2020stateful} to this configuration does not increase (and in many cases reduces) the ASR.

Column 8 then presents our full \name attack configuration including both \textit{adapt} and \textit{resample} mechanisms that perform rejection sampling with proposal fine-tuning. All attacks show significant gains in ASR, with most attacks exceeding 80\% (and in many cases 90\%) ASR across all datasets. This column is the first to propose attacks that are widely successful in attacking existing SDMs, and suggests that both \textit{adapt} and \textit{resample} mechanisms are important for successful attacks.

\paragraph{\textbf{OSD}}
As discussed in Section~\ref{sec:att_extract}, if we can determine with a simple diagnostic test if the SDM in place is banning accounts and using a per-account query store, it may be more effective to simply run the standard attack under multiple (but relatively few) accounts. 

We report the cost and ASR of what it would take to run the diagnostic tests in Section~\ref{sec:att_extract} then run the standard attacks. We find that we can compute these attacks in under 25 accounts, with Square~\cite{andriushchenko2020square} only requiring 2 and SurFree~\cite{maho2021surfree} requiring 3, even with the diagnostic test. Note that by using the first query used in the attack algorithm as the point $x$ for the diagnostic test, we can reuse the query on account $B$ to proceed with the attack, with either the standard or the \name version.

\begin{table}[]
\centering
\small
\caption{We can attack OSD with few accounts by running our diagnostic tests to determine the defense is account banning based and then running the standard attacks. Results are presented in (ASR / query count) format. The cost of the diagnostic test ($\sim$ 50 extra queries, 1 extra account) is included in the reported numbers. Square and SurFree can be completed in an average of 2 and 3 accounts respectively.}
\begin{tabular}{lcc}
\toprule
\multirow{2}{*}{\textbf{Attack}}       & \multicolumn{2}{c}{\textbf{OSD}}                     \\ \cmidrule(lr){2-3} 
                                       & \multicolumn{1}{c}{\textbf{ASR}} & \textbf{\# Accounts}  \\\midrule 
\textbf{NES}                           & \multicolumn{1}{c}{100\% / 861}  & 18                    \\ 
\textbf{Square}                        & \multicolumn{1}{c}{98\% / 67}             &   2                    \\ 
\textbf{HSJA}                          & \multicolumn{1}{c}{100\% / 1063}             &                21       \\
\textbf{QEBA}                          & \multicolumn{1}{c}{100\% / 948}             &              19         \\ 
\multicolumn{1}{l}{\textbf{SurFree}} & \multicolumn{1}{c}{100\% / 146}             &  \multicolumn{1}{c}{3} \\ 
\textbf{Boundary}                      &  \multicolumn{1}{c}{100\% / 680}             &   14                    \\ \bottomrule
\end{tabular}
\end{table}

\begin{figure}
    \centering
    \includegraphics[width=\columnwidth]{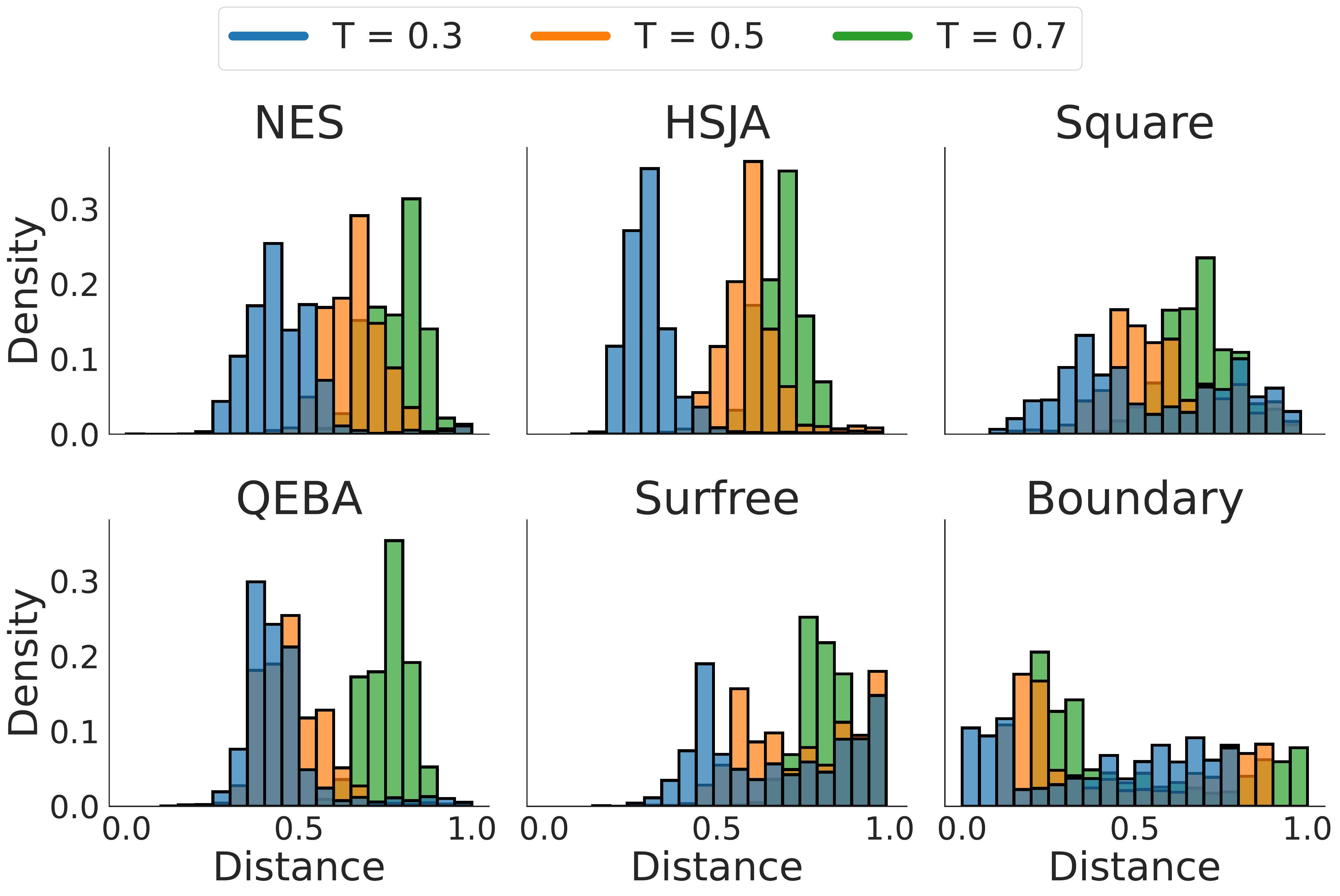}
    \caption{Our attacks automatically increase the distance between queries in response to the defense increasing its collision threshold. The horizontal axis shows pairwise distance between attack queries measured using the defense's distance function. Each histogram (blue, orange, green) corresponds to attacking an increasingly larger collision threshold used by the defense. Attacks are launched using the CIFAR10 dataset against Blacklight with a collision threshold  $\in \{0.3,0.5,0.7\}$}
    \label{fig:threshold_similarity}
\end{figure}

\subsection{Our Adaptive Attacks vs. Reconfigured SDMs}
\label{sec:reconfigs_sdm}

We now explore if SDMs can be easily reconfigured to be robust against \name-enhanced attacks. Below we focus on Blacklight~\cite{li2022blacklight} as the most recent defense where we consider two reconfigurations that target its two key mechanisms: the similarity procedure and feature extraction.

\paragraph{\textbf{Increasing the detection threshold.}} As our adaptive attacks use less similar queries, we evaluate raising the SDM's similarity threshold. This change should increase attack detection rate as more queries would collide. However, increasing Blacklight's similarity threshold drastically impacts its false positive rate, hurting its natural performance. Specifically, raising the threshold from the default 0.5 to 0.7 more than quadruples the false positive rate (from 0.2\% to 0.9\%) on $\approx$1 million images sampled from the tiny images dataset, which is the original database from which CIFAR10 is sampled~\cite{krizhevsky2009learning}. Furthermore, even if the defense is willing to make this sacrifice, the ASR remains quite high ($>63\%$ for all attacks). As an explanation, we plot the average pairwise distances between queries per threshold in Figure~\ref{fig:threshold_similarity}. The figure showcases the adaptive nature of our attacks: they automatically tune the distance between queries given the SDM settings. This observation follows from the design of the attacks that use rejection sampling. Further, our results suggest that natural performance drops off before adaptive attack is forced to use disruptive enough queries.

\paragraph{\textbf{Adjusting the feature extractor's hyperparameters.}}
We investigate whether changing Blacklight's~\cite{li2022blacklight} feature extractor hyperparameters improves robustness to our adaptive attacks. We change the window size and quantization step size for Blacklight, and set the threshold to meet the default 0.2\% false positive rate. The results are shown in Table~\ref{tab:alt_configs}. As expected, we find that each of the tested settings are still vulnerable to our adaptive attack, with each setting susceptible to an attack that achieves 99\% ASR. Attacks under all alternative Blacklight configurations achieve a 80\% or greater ASR. These results suggest that creating truly robust SDMs is a challenging problem and will require more than simple tweaks of existing SDMs. 

\begin{table}[t]
\small
\centering
\caption{Our proposed adaptive attacks (adapt and resample mechanisms) continue to adapt and succeed with high ASR against SDMs even with adjusted hyperaparameter settings. Results are presented in (ASR / query count) format. Attacks are launched using the CIFAR10 dataset against Blacklight with 5 different configurations (1 default + 4 variations that adjust window size $w$ and quantization step size $q$). The column headers represent ($w$,$q$).}\label{tab:alt_configs}
\begin{tabular}{l@{\hspace{1.5mm}} c@{\hspace{1.5mm}}  c@{\hspace{1.5mm}} c@{\hspace{1.5mm}} c@{\hspace{1.5mm}} c@{\hspace{1.5mm}}}
\toprule
\multicolumn{1}{l}{\multirow{2}{*}{\textbf{Attack}}} & \multicolumn{5}{c}{\textbf{Blacklight Alternate Configurations}}                                                                                                                           \\ \cmidrule{2-6} 
\multicolumn{1}{l}{}                                 & {\textbf{(50,20)}} & {\textbf{(20,20)}} & {\textbf{(100,20)}} & {\textbf{(50,10)}} & \textbf{(50,50)} \\ \hline
\textbf{ NES}                                           & {99\% / 1540}        & {97\% / 1548}                   & {97\% / 1548}                    & {96\% / 1576}                   &      98\% / 1585              \\
\textbf{ Square}                                        & {93\% / 218}         & {93\% / 193}                   & {94\% / 187}                    & {93\% / 193}                   &      97\% / 184              \\
\textbf{ HSJA}                                          & {82\% / 1615}        & {88\% / 1636}                   & {89\% / 1786}                    & {85\% / 1721}                   &      91\% / 1735              \\
\textbf{ QEBA}                                          & {86\% / 1627}        & {100\% / 1047}                   & {96\% / 1269}                    & {95\% / 1405}                   &      100\% / 1009              \\ 
\textbf{ SurFree}                                       & {81\% / 145}         & {99\% / 186}                   & {81\% / 144}                    & {80\% / 142}                   &    {97\% / 189}                \\ 
\textbf{ Bound.}                                      & {98\% / 3302}            & {94\% / 1810}                   & {99\% / 3302}                    & {99\% / 3317}                   &      89\% / 1785              \\ \bottomrule
\end{tabular}
\end{table}

\subsection{Attack Cost}
\label{sec:eval_cost}
We report the average number of queries needed by successful attacks in Table~\ref{tab:exp_global}. In most cases, \name attacks complete in less than $10$k queries, with a few exceptions. Compared to attacks against an undefended model, the SDM can indeed raise the number of queries. For example, NES (score based; targeted) against CIFAR10 typically requires $\sim900$ queries for an undefended model, while \name-NES takes $\sim1.5$k queries against a Blacklight-defended model. 
However, modern MLaaS platforms such as Clarifai, Google, and Microsoft, typically charge only \$1-\$1.50 per $1$k queries. In such a case, the attacker would only pay an additional \$<1 to successfully generate an adversarial example against a defended model. This overhead depends on the threat model (soft/hard label; targeted/un-targeted) and varies among the different attacks. We observe the largest overhead for the weakest threat model (hard label; targeted) where a HSJA attacker would need to pay an additional $\sim\$18$ to attack a ImageNet model.

Note that standard attacks are both expensive and perform poorly against SDM-defended models.  Assuming a typical $100k$ query budget~\cite{li2022blacklight} they would cost $>\$100$ and typically fail to find adversarial examples against recent SDM models.
\section{Discussion}\label{sec:discussion}
We now discuss possible SDM countermeasures to our \name attacks, the applicability of our \name attacks to other domains such as text, additional considerations on attacking per-account SDMs, limitations, and ethical considerations.

\subsection{Countermeasures}\label{sec:disc_counter}
While the majority of this work focused on understanding and evaluating existing SDMS (and their reconfigurations), we now consider alternative configurations. New SDMs can be designed by considering alternate choices for the feature extractor and action module. Against our proposed adaptive attacks, would alternate SDM designs work better?

\paragraph{\textbf{Alternative Feature Extractors:}} A vast body of work exists on image feature extractors, outside the realm of adversarial defense~\cite{du2020perceptual}. We provide preliminary insight into understanding their viability as SDM feature extractors by evaluating the robust accuracy of SDM configurations with two popular perceptual feature extractors: PHash and Facebook's PDQ~\cite{dalins2019pdq}. Specifically, we take Blacklight and only replace its feature extractor with PHash and PDQ.  We find that these hash functions do not provide any gains in robustness, e.g., \name-NES attacks are able to achieve 99\% and 96\% ASR respectively. This suggests that selecting an ideal feature extractor for SDMs is still a challenging problem for future work.

\paragraph{\textbf{Alternative Action Function:}} Our adaptive attacks use multiple queries to extract information about the SDM. This is possible because the SDM itself leaks information while accepting/rejecting queries. One way for SDMs to avoid this is to respond with random labels or noisy probability scores. Such an action function could make it harder for an adversary to adapt. However, even so, an adaptive attack could estimate whether a response is random or not by analyzing the response distribution. Moreover, returning random responses could lead to unusual behaviour, i.e., querying the same image twice in a row against an SDM with random label responses could lead to a ``free'' adversarial example (with a perturbation $\epsilon=0$).

\paragraph{\textbf{Adaptive SDMs with Ensembling:}} One alternative for the defender is to try ensembling SDMs, in the hope that they can collectively prevent attacks. However, \name does not make assumptions about the underlying defense and will treat the entire ensemble as a single black-box model, and should thus continue to be effective. We confirm this observation by evaluating vanilla and \name attacks against an ensemble of Blacklight and PIHA on CIFAR10 in Table~\ref{tab:ensemble}. We find that our attacks (without additional modification) are still successful, e.g., \name-Square achieves an ASR of 95\%. Furthermore, ensembling SDMs would raise the false positive rate of the system, suggesting that such an approach requires deeper investigation to be practical.

\begin{table}[t]
\caption{\name attacks continue to be effective against ensembles of multiple SDMs. Results are presented in (ASR / query count) format. Attacks are launched using the CIFAR10 dataset against an ensemble of Blacklight and PIHA.}\label{tab:ensemble}
\begin{tabular}{@{}lcc@{}}
\toprule
\textbf{Attack}   & \textbf{Standard} & \textbf{Adapt + Resample (\name)} \\ \midrule
\textbf{NES}      & 0\%               &   81\% / 1627                        \\
\textbf{Square}   & 0\%              &   95\% / 205                        \\
\textbf{HSJA}     & 0\%               &   67\% / 3159                        \\
\textbf{QEBA}     & 0\%               &   91\% / 2078                        \\
\textbf{SurFree}  & 0\%               &   63\% / 137                        \\
\textbf{Boundary} & 0\%               &   89\% / 2821                        \\ \bottomrule

\end{tabular}
\end{table}

\subsection{Applicability to Other Domains}
Some SDMs claim to defend well against adversarial attacks in other domains such as text classification. For example, Blacklight claims perfect detection of the TextFooler synonym-subsitution adversarial attack on the IMDB dataset. TextFooler proceeds by (a) removing one word at a time and querying to identify an importance ordering, and (b) querying a series of synonym substitutions for the important words to elicit misclassification. To defend against this attack, Blacklight converts an input text to its embedding representation, and uses it to compute a similarity hash (similar to the image domain). We are able to confirm this observation --- TextFooler's ASR is reduced from 100\% to 0\% against Blacklight on a BERT classifier for a test set of 100 samples. 

Given the success of the \name strategy in the image domain, one might consider its viability in other domains such as text. To this end, we now consider some preliminary results to show that it may indeed be applicable here as well. Given text sample $x$, we model the proposal distribution for all ``similar'' queries that also do no collide with $x$ as the parametric edit-distance based sentence distribution. This distribution represents all sentences that are at some Levenshtein distance $\lambda$ from $x$. Then, proposal $p_\lambda$ can be adapted by performing oracle guided fine-tuning of $\lambda$. Finally, one can sample from fine-tuned distribution $p_\lambda$ to construct attack queries --- in practice, this amounts to insertion/removal of spaces and words from a query until edit distance is $\lambda$. We find that this approach is able to achieve a 78\% ASR with an average of 20\% words perturbed. This suggests that the \name approach is likely relevant for other domains --- we leave improvement of the ASR/reduction of the perturbed word count to future work.

\subsection{Per-Account SDMs}
One consideration that might make OSD~\cite{chen2020stateful} more suitable is a case where the assumption that accounts are easy to create in large quantities (through Sybil accounts~\cite{douceur2002sybil,yang2014uncovering}, for example) is no longer valid. However, we point out that due to the existence of improved attacks since the original paper (such as SurFree~\cite{maho2021surfree}), attacks can now be completed in as few as 2-3 accounts limiting this concern.

Another possibility is to consider reducing $k$ (i.e., the number of queries required to cause an account ban). However, OSD's original analysis~\cite{chen2020stateful} suggests that this cannot be done without drastically sacrificing the false positive rate. In the future, if some defense that can use a lower $k$ can be created, one can simply use the diagnostic tests described in Section~\ref{sec:att_extract} to compare the expected attack query costs of the attack and the expected overhead costs of \name against the number of ``free'' queries per account and decide which strategy is better.

\subsection{Generalizability}
We note that the \name approach of adapting proposal distributions and resampling substitute queries upon query collisions is general beyond the specific attack algorithms evaluated in the paper. \name accommodates black-box attacks that comprise the three common elements discussed in Section~\ref{sec:bb_elements}. In Section~\ref{sec:att_existing}, we show how to apply \name to six black-box attacks that differ considerably (for example, NES~\cite{ilyas2018black} relies on finite differences to estimate the gradient, whereas Square performs random search along square perturbations). These six attacks thus provide blueprints for applying \name to a wide range of new attacks. Specifically, one should start with identifying the pieces of the attack that map to the three common elements in Section~\ref{sec:bb_elements} and apply the corresponding modifications to adapt the proposal distribution and resample as applicable. Note also that \name is agnostic to the defense deployed, and should apply to future SDMs that leak information (See Section~\ref{sec:disc_counter}).

\subsection{Overhead of \name}
Section~\ref{sec:eval_cost} discusses the query overhead when running \name. In general, this query overhead can depend upon two factors. First, the general trend suggests that more classes and higher input dimensionality incur a higher cost (this is also typical for standard attacks). Second, the query overhead appears to depend on the SDM deployed; we likely need more SDMs in existence to draw firmer conclusions on the exact relationship.

Another minor overhead is selecting the hyperparameters for the adaptation of the proposal distribution (e.g., upper/lower limits for adapting the variance of the Gaussian distribution). We emphasize that \name uses the same hyperparameters regardless of the SDM (since the SDM is unknown). The hyperparameters only vary across datasets, which is consistent with standard black-box attacks. These attacks select hyperparameters that are dataset-specific to handle the different convergence rates based on the input dimensionality. Given an attack/dataset, we found it fairly simple to select hyperparameters by initializing them conservatively and observing convergence over 1-2 samples. Such hyperparameter selection typically took fewer than 15 minutes of experimentation and about 10 iterations of around 100 queries each, or about 1000 queries overall. This is a one-time cost per dataset.

\subsection{Limitations}
There are a few related attacks and defenses we do not consider in this work. Firstly, we do not consider transfer attacks. Although transfer attacks can be crafted with the assistance of techniques such as model stealing~\cite{juuti2019prada}, we follow prior work on SDMs and consider transfer attacks to be an orthogonal problem --- transfer attack defenses already exist~\cite{tramer2017ensemble} and can be combined with stateful systems to build a more complete defense. 

Secondly, we do not consider alternative approaches to detecting black-box attacks such as Adversarial Attack on Attackers (AAA)~\cite{chen2022adversarial} and AdvMind~\cite{pang2020advmind}. AAA is a post-processing attack that attempts to modify the logits loss curve to locally point in the incorrect attack direction in a periodic fashion, misguiding score-based query attacks from a successful attack. AdvMind is a detection model that infers the intent of an adversary and detect attacks. While these approaches both aim to thwart black-box attacks, they are orthogonal to SDMs that aim to detect similar queries. 

\subsection{Ethical Considerations}
While our paper exposes new attack strategies that could then be used to attack real-world systems deploying SDMs, it is important for developers to have the system, software, and algorithmic tools necessary to truly understand the potential vulnerabilities and true robustness of any SDM that may be in consideration for future deployment. We note that query-based black-box attacks already exist and have been used to attack real-world systems already~\cite{li2020qeba}. Our hope is that, analogously to the white-box setting, our work encourages stronger evaluation of SDMs before we reach a point where they are being falsely relied on to solve the robustness problem in real-world deployments.
\section{Conclusion}
This paper proposes \name, an adaptive black-box attack strategy that significantly increases the attack success rate against four state-of-the-art SDMs. Our key insight is that SDMs leak information about the similarity-detection procedure and its parameters, enabling an attacker to adapt its queries to evade collisions. Our work shows that these SDMs are not as truly robust as previously believed, and provides a new benchmark to test future SDMs.

\section*{Acknowledgements}
This material is based upon work supported by DARPA
under agreement number 885000, National Science Foundation Grant No. 2039445, and National Science Foundation
Graduate Research Fellowship Grant No. DGE 1841052.
Any opinion, findings, and conclusions or recommendations
expressed in this material are those of the authors(s) and do
not necessarily reflect the views of our research sponsors.

\begin{table*}[!b]
\resizebox{0.8\textwidth}{!}{%
\begin{tabular}{@{}lllllllrllll@{}}
\toprule
\multirow{2}{*}{\textbf{Dataset}}                               &                    \multirow{2}{*}{\textbf{Attack}}                    & \multirow{2}{*}{$ge_{tries}$}     & \multicolumn{4}{c}{$\sigma$} &   \multirow{2}{*}{$steps_{tries}$} & \multicolumn{4}{c}{$step$}                              \\ \cmidrule{4-7} \cmidrule{9-12}
& & & $stps$ & $sam$ & $cr$ & $[\sigma_{lo}$, $\sigma_{hi}]$ & & $stps$ & $sam$ & $cr$ & $[step_{lo}$, $step_{hi}]$\\ \midrule
\multicolumn{1}{l|}{\multirow{6}{*}{\textbf{CIFAR}}}    & \multicolumn{1}{|l|}{\textbf{NES}}      & \multicolumn{1}{c}{5}  & \multicolumn{1}{c}{10} & \multicolumn{1}{c}{20} & \multicolumn{1}{c}{0}    & \multicolumn{1}{c}{{[}0.05, 0.5{]}}   & \multicolumn{1}{c}{20}                 & \multicolumn{1}{c}{10}                              & \multicolumn{1}{c}{20}                                  & \multicolumn{1}{c}{0.005}                                & \multicolumn{1}{c}{{[}0.1,0.1{]}}            \\  
         & \multicolumn{1}{|l|}{\textbf{Square}}   & \multicolumn{1}{c}{-}  & \multicolumn{1}{c}{-}  & \multicolumn{1}{c}{-}  & \multicolumn{1}{c}{-}    & \multicolumn{1}{c}{-}                 & \multicolumn{1}{c}{300}                & \multicolumn{1}{c}{10}                              & \multicolumn{1}{c}{5   10 } & \multicolumn{1}{c}{0   0.5 } & \multicolumn{1}{c}{{[}10,100{]}, {[}1,32{]}} \\  
         & \multicolumn{1}{|l|}{\textbf{HSJA}}     & \multicolumn{1}{c}{20} & \multicolumn{1}{c}{10} & \multicolumn{1}{c}{20} & \multicolumn{1}{c}{0.05} & \multicolumn{1}{c}{{[}1,2{]}}         & \multicolumn{1}{c}{5}                  & \multicolumn{1}{c}{-}                               & \multicolumn{1}{c}{-}                                   & \multicolumn{1}{c}{-}                                    & \multicolumn{1}{c}{-}                        \\  
         & \multicolumn{1}{|l|}{\textbf{QEBA}}     & \multicolumn{1}{c}{20} & \multicolumn{1}{c}{10} & \multicolumn{1}{c}{20} & \multicolumn{1}{c}{0.05} & \multicolumn{1}{c}{{[}1,2{]}}         & \multicolumn{1}{c}{5}                  & \multicolumn{1}{c}{-}                               & \multicolumn{1}{c}{-}                                   & \multicolumn{1}{c}{-}                                    & \multicolumn{1}{c}{-}                        \\  
         & \multicolumn{1}{|l|}{\textbf{SurFree}}  & \multicolumn{1}{c}{-}  & \multicolumn{1}{c}{-}  & \multicolumn{1}{c}{-}  & \multicolumn{1}{c}{-}    & \multicolumn{1}{c}{-}                 & \multicolumn{1}{c}{100}                & \multicolumn{1}{c}{5}                               & \multicolumn{1}{c}{20}                                  & \multicolumn{1}{c}{0.05}                                 & \multicolumn{1}{c}{{[}5,50{]}}               \\  
         & \multicolumn{1}{|l|}{\textbf{Boundary}} & \multicolumn{1}{c}{-}   & \multicolumn{1}{c}{-}   & \multicolumn{1}{c}{-}   & \multicolumn{1}{c}{-}     & \multicolumn{1}{c}{-}                  & \multicolumn{1}{c}{-}                   & \multicolumn{1}{c}{-}                                & \multicolumn{1}{c}{-}                                    & \multicolumn{1}{c}{-}                                     & \multicolumn{1}{c}{-}                         \\  
\midrule
\multirow{6}{*}{\textbf{ImageNet}} & \multicolumn{1}{|l|}{\textbf{NES}}      & \multicolumn{1}{c}{5}  & \multicolumn{1}{c}{10} & \multicolumn{1}{c}{20} & \multicolumn{1}{c}{0.05} & \multicolumn{1}{c}{{[}0.01, 0.5{]}}   & \multicolumn{1}{c}{5}                  & \multicolumn{1}{c}{10}                              & \multicolumn{1}{c}{20}                                  & \multicolumn{1}{c}{0.13}                                 & \multicolumn{1}{c}{{[}0.00005, 0.1{]}}       \\  
         & \multicolumn{1}{|l|}{\textbf{Square}}   & \multicolumn{1}{c}{-}  & \multicolumn{1}{c}{-}  & \multicolumn{1}{c}{-}  & \multicolumn{1}{c}{-}    & \multicolumn{1}{c}{-}                 & \multicolumn{1}{c}{300}                & \multicolumn{1}{c}{3 10} & \multicolumn{1}{c}{5   10 } & \multicolumn{1}{c}{0   0.5 } & \multicolumn{1}{c}{{[}50,100{]},{[}1,100{]}} \\  
         & \multicolumn{1}{|l|}{\textbf{HSJA}}     & \multicolumn{1}{c}{20} & \multicolumn{1}{c}{10} & \multicolumn{1}{c}{20} & \multicolumn{1}{c}{0.05} & \multicolumn{1}{c}{{[}1.5,5{]}}       & \multicolumn{1}{c}{5}                  & \multicolumn{1}{c}{-}                               & \multicolumn{1}{c}{-}                                   & \multicolumn{1}{c}{-}                                    & \multicolumn{1}{c}{-}                        \\  
         & \multicolumn{1}{|l|}{\textbf{QEBA}}     & \multicolumn{1}{c}{20} & \multicolumn{1}{c}{10} & \multicolumn{1}{c}{20} & \multicolumn{1}{c}{0.05} & \multicolumn{1}{c}{{[}1.5,5{]}}       & \multicolumn{1}{c}{5}                  & \multicolumn{1}{c}{-}                               & \multicolumn{1}{c}{-}                                   & \multicolumn{1}{c}{-}                                    & \multicolumn{1}{c}{-}                        \\  
         & \multicolumn{1}{|l|}{\textbf{SurFree}}  & \multicolumn{1}{c}{-}  & \multicolumn{1}{c}{-}  & \multicolumn{1}{c}{-}  & \multicolumn{1}{c}{-}    & \multicolumn{1}{c}{-}                 & \multicolumn{1}{c}{100}                & \multicolumn{1}{c}{5}                               & \multicolumn{1}{c}{20}                                  & \multicolumn{1}{c}{0.05}                                 & \multicolumn{1}{c}{{[}5,50{]}}               \\  
         & \multicolumn{1}{|l|}{\textbf{Boundary}} & \multicolumn{1}{c}{-}   & \multicolumn{1}{c}{-}   & \multicolumn{1}{c}{-}   & \multicolumn{1}{c}{-}     & \multicolumn{1}{c}{-}                  & \multicolumn{1}{c}{-}                   & \multicolumn{1}{c}{-}                                & \multicolumn{1}{c}{-}                                    & \multicolumn{1}{c}{-}                                     & \multicolumn{1}{c}{-}                         \\  
\midrule 
\multirow{6}{*}{\textbf{CelebaHQ}} & \multicolumn{1}{|l|}{\textbf{NES}}      & \multicolumn{1}{c}{5}  & \multicolumn{1}{c}{10} & \multicolumn{1}{c}{20} & \multicolumn{1}{c}{0.05} & \multicolumn{1}{c}{{[}0.01, 0.5{]}}   & \multicolumn{1}{c}{5}                  & \multicolumn{1}{c}{10}                              & \multicolumn{1}{c}{20}                                  & \multicolumn{1}{c}{0.13}                                 & \multicolumn{1}{c}{{[}0.00005, 0.1{]}}       \\  
         & \multicolumn{1}{|l|}{\textbf{Square}}   & \multicolumn{1}{c}{-}  & \multicolumn{1}{c}{-}  & \multicolumn{1}{c}{-}  & \multicolumn{1}{c}{-}    & \multicolumn{1}{c}{-}                 & \multicolumn{1}{c}{300}                & \multicolumn{1}{c}{3 10} & \multicolumn{1}{c}{5   10 } & \multicolumn{1}{c}{0   0.5 } & \multicolumn{1}{c}{{[}50,100{]},{[}1,100{]}} \\  
         & \multicolumn{1}{|l|}{\textbf{HSJA}}     & \multicolumn{1}{c}{20} & \multicolumn{1}{c}{10} & \multicolumn{1}{c}{20} & \multicolumn{1}{c}{0.05} & \multicolumn{1}{c}{{[}0.5,5{]}}       & \multicolumn{1}{c}{5}                  & \multicolumn{1}{c}{-}                               & \multicolumn{1}{c}{-}                                   & \multicolumn{1}{c}{-}                                    & \multicolumn{1}{c}{-}                        \\  
         & \multicolumn{1}{|l|}{\textbf{QEBA}}     & \multicolumn{1}{c}{20} & \multicolumn{1}{c}{10} & \multicolumn{1}{c}{20} & \multicolumn{1}{c}{0.05} & \multicolumn{1}{c}{{[}0.5,5{]}}       & \multicolumn{1}{c}{5}                  & \multicolumn{1}{c}{-}                               & \multicolumn{1}{c}{-}                                   & \multicolumn{1}{c}{-}                                    & \multicolumn{1}{c}{-}                        \\  
         & \multicolumn{1}{|l|}{\textbf{SurFree}}  & \multicolumn{1}{c}{-}  & \multicolumn{1}{c}{-}  & \multicolumn{1}{c}{-}  & \multicolumn{1}{c}{-}    & \multicolumn{1}{c}{-}                 & \multicolumn{1}{c}{100}                & \multicolumn{1}{c}{5}                               & \multicolumn{1}{c}{20}                                  & \multicolumn{1}{c}{0.05}                                 & \multicolumn{1}{c}{{[}5,50{]}}               \\  
         & \multicolumn{1}{|l|}{\textbf{Boundary}} & \multicolumn{1}{c}{-}   & \multicolumn{1}{c}{-}   & \multicolumn{1}{c}{-}   & \multicolumn{1}{c}{-}     & \multicolumn{1}{c}{-}                  & \multicolumn{1}{c}{-}                   & \multicolumn{1}{c}{-}                                & \multicolumn{1}{c}{-}                                    & \multicolumn{1}{c}{-}                                     & \multicolumn{1}{c}{-}                         \\  
 \midrule
\multirow{6}{*}{\textbf{IIoT}}     & \multicolumn{1}{|l|}{\textbf{NES}}      & \multicolumn{1}{c}{5}  & \multicolumn{1}{c}{10} & \multicolumn{1}{c}{20} & \multicolumn{1}{c}{0.1}  & \multicolumn{1}{c}{{[}0.00001,0.5{]}} & \multicolumn{1}{c}{5}                  & \multicolumn{1}{c}{10}                              & \multicolumn{1}{c}{20}                                  & \multicolumn{1}{c}{0.05}                                 & \multicolumn{1}{c}{{[}0.00005,1{]}}          \\  
         & \multicolumn{1}{|l|}{\textbf{Square}}   & \multicolumn{1}{c}{-}  & \multicolumn{1}{c}{-}  & \multicolumn{1}{c}{-}  & \multicolumn{1}{c}{-}    & \multicolumn{1}{c}{-}                 & \multicolumn{1}{c}{5}                  & \multicolumn{1}{c}{10}                              & \multicolumn{1}{c}{5   10 } & \multicolumn{1}{c}{0   0.5 } & \multicolumn{1}{c}{{[}10,100{]}, {[}1,32{]}} \\  
         & \multicolumn{1}{|l|}{\textbf{HSJA}}     & \multicolumn{1}{c}{1}  & \multicolumn{1}{c}{20} & \multicolumn{1}{c}{20} & \multicolumn{1}{c}{0.05} & \multicolumn{1}{c}{{[}0.001,20{]}}    & \multicolumn{1}{c}{5}                  & \multicolumn{1}{c}{20}                              & \multicolumn{1}{c}{20}                                  & \multicolumn{1}{c}{0.05}                                 & \multicolumn{1}{c}{{[}0.00005,1{]}}          \\  
         & \multicolumn{1}{|l|}{\textbf{QEBA}}     & \multicolumn{1}{c}{1}  & \multicolumn{1}{c}{20} & \multicolumn{1}{c}{20} & \multicolumn{1}{c}{0.05} & \multicolumn{1}{c}{{[}0.001,20{]}}    & \multicolumn{1}{c}{5}                  & \multicolumn{1}{c}{20}                              & \multicolumn{1}{c}{20}                                  & \multicolumn{1}{c}{0.05}                                 & \multicolumn{1}{c}{{[}0.00005,1{]}}          \\  
         & \multicolumn{1}{|l|}{\textbf{SurFree}}  & \multicolumn{1}{c}{-}  & \multicolumn{1}{c}{-}  & \multicolumn{1}{c}{-}  & \multicolumn{1}{c}{-}    & \multicolumn{1}{c}{-}                 & \multicolumn{1}{c}{10}                 & \multicolumn{1}{c}{10}                              & \multicolumn{1}{c}{20}                                  & \multicolumn{1}{c}{0.05}                                 & \multicolumn{1}{c}{{[}1,50{]}}               \\  
                               & \multicolumn{1}{|l|}{\textbf{Boundary}}                      &    \multicolumn{1}{c}{-}                     &     \multicolumn{1}{c}{-}                    &    \multicolumn{1}{c}{-}                     &    \multicolumn{1}{c}{-}                       &    \multicolumn{1}{c}{-}                                    & \multicolumn{1}{c}{-}                    &      \multicolumn{1}{c}{-}                                                &    \multicolumn{1}{c}{-}                                                      &  \multicolumn{1}{c}{-}                                                         & \multicolumn{1}{c}{-}                                               \\ \bottomrule
\end{tabular}
}
\caption{\name adapt and resample hyperparameters for all attacks and datasets. Hyperparameters are selected on a per-attack basis, i.e., $lo$ and $hi$ values by observing valid ranges of values that typically work with the standard attack on undefended models, and $stps$, $sam$ chosen to obtain a stable estimate of $cr$ which is generally set to be low.}
\label{tab:adaptive_hyperparams}
\end{table*}

\bibliographystyle{ACM-Reference-Format}
\bibliography{sample}

\appendix
\appendix

\section{Additional Evaluation}\label{sec:app_add_eval}
As discussed in Section~\ref{sec:eval_vanilla}, Table~\ref{tab:cifar1000} presents extended results of \name against a larger sample of 1000 CIFAR10 images against Blacklight.
\section{Attack Hyperparameters}\label{sec:app_hyper}
As discussed in Section~\ref{sec:eval_vanilla}, Table~\ref{tab:adaptive_hyperparams} presents hyperparameters for \name adapt and resample attacks presented in Table~\ref{tab:exp_global}.

\begin{table}[h]
\small
\caption{Extended results for \name attacks with adapt and resample. Results are computed over 1000 images on CIFAR10 against Blacklight.}\label{tab:cifar1000}
\begin{tabular}{@{}lcc@{}}
\toprule
\textbf{Attack}   & \textbf{Standard} & \textbf{Adapt + Resample} \\ \midrule
\textbf{NES}      & 0\%               &   96\% / 1585                        \\
\textbf{Square}   & 0\%               &   93\% / 206                        \\
\textbf{HSJA}     & 0\%               &   86\% / 1573                        \\
\textbf{QEBA}     & 0\%               &   97\% / 1324                        \\
\textbf{SurFree}  & 0\%               &   83\% / 149                        \\
\textbf{Boundary} & 0\%               &   99\% / 2819                        \\ \bottomrule
\end{tabular}
\end{table}

\end{document}